\documentclass[aps,prr,twocolumn,superscriptaddress]{revtex4-1}  
\usepackage{graphicx}  
\usepackage{dcolumn}   
\usepackage{bm}        
\usepackage{amssymb}   
\usepackage{siunitx}
\usepackage{amsmath}
\usepackage{float}
\usepackage{prettyref}
\usepackage{filecontents}
\usepackage[dvipsnames]{xcolor}
\usepackage[colorlinks,linkcolor=blue,citecolor=blue,urlcolor=blue]{hyperref}
\usepackage{textgreek}

\newrefformat{fig}{Fig.~\ref{#1}}

\newcommand{\IcRn}{\ensuremath{{I}_{{\textrm{C}}}{R}_{\textrm{N}}}}
\newcommand{\IexRn}{\ensuremath{{I}_{{\textrm{ex}}}{R}_{\textrm{N}}}}

\newcommand{\Del}{\ensuremath{\mathit{\Delta}_{{\textrm{el}}}}}
\newcommand{\SSm}{S-Sm}
\newcommand{\SSSm}{S'-S-Sm}
\newcommand{\Is}{\ensuremath{{I}_{{\textrm{S}}}}}

\newcommand{\Ic}{\ensuremath{{I}_{{\textrm{C}}}}}

\newcommand{\Vg}{\ensuremath{{V}_{\textrm{G}}}}

\newcommand{\meanfreepath}{\ensuremath{ {l}_{\textrm{e}}}}

\begin{document}



\title{Enhancement of Proximity-Induced Superconductivity in a Planar Ge Hole Gas}

\author{Kushagra Aggarwal}
\email{kushagra.aggarwal@ist.ac.at}
\affiliation{Institute of Science and Technology Austria, Am Campus 1, 3400 Klosterneuburg, Austria}
\author{Andrea Hofmann}
\affiliation{Institute of Science and Technology Austria, Am Campus 1, 3400 Klosterneuburg, Austria}
\author{Daniel Jirovec}
\affiliation{Institute of Science and Technology Austria, Am Campus 1, 3400 Klosterneuburg, Austria}
\author{Ivan Prieto}
\affiliation{Institute of Science and Technology Austria, Am Campus 1, 3400 Klosterneuburg, Austria}
\author{Amir Sammak}
\affiliation{QuTech and Netherlands Organisation for Applied Scientific Research (TNO), Stieltjesweg 1, 2628 CK Delft, The Netherlands}
\author{Marc Botifoll}
\affiliation{Catalan Institute of Nanoscience and Nanotechnology (ICN2), CSIC and BIST, Campus UAB, 08193 Bellaterra, Barcelona, Catalonia, Spain}
\author{Sara Mart\'i-S\'anchez}
\affiliation{Catalan Institute of Nanoscience and Nanotechnology (ICN2), CSIC and BIST, Campus UAB, 08193 Bellaterra, Barcelona, Catalonia, Spain}
\author{Menno Veldhorst}
\affiliation{QuTech and Kavli Institute of Nanoscience, Delft University of Technology, Lorentzweg 1, 2628 CJ Delft, The Netherlands}
\author{Jordi Arbiol}
\affiliation{Catalan Institute of Nanoscience and Nanotechnology (ICN2), CSIC and BIST, Campus UAB, 08193 Bellaterra, Barcelona, Catalonia, Spain}
\affiliation{ICREA, Pg. Lluís Companys 23, 08010 Barcelona, Catalonia, Spain}
\author{Giordano Scappucci}
\affiliation{QuTech and Kavli Institute of Nanoscience, Delft University of Technology, Lorentzweg 1, 2628 CJ Delft, The Netherlands}
\author{Jeroen Danon}
\affiliation{Center for Quantum Spintronics, Department of Physics, Norwegian University of Science and Technology, NO-7491 Trondheim, Norway}
\author{Georgios Katsaros}
\email{georgios.katsaros@ist.ac.at}
\affiliation{Institute of Science and Technology Austria, Am Campus 1, 3400 Klosterneuburg, Austria}

\date{\today}

\begin{abstract}
	Hole gases in planar germanium can have high mobilities in combination with strong spin-orbit interaction and electrically tunable g-factors, and are therefore emerging as a promising platform for creating hybrid superconductor-semiconductor devices.
	A key challenge towards hybrid Ge-based quantum technologies is the design of high-quality interfaces and superconducting contacts that are robust against magnetic fields.
	In this work, by combining the assets of aluminum, which provides good contact to the Ge, and niobium, which has a significant superconducting gap, we demonstrate highly transparent low-disordered JoFETs with relatively large \IcRn \ products that are capable of withstanding high magnetic fields.
	We furthermore demonstrate the ability of phase-biasing individual JoFETs, opening up an avenue to explore topological superconductivity in planar Ge.
	The persistence of superconductivity in the reported hybrid devices beyond 1.8 Tesla paves the way towards integrating spin qubits and proximity-induced superconductivity on the same chip.
\end{abstract}

\maketitle
\section{Introduction}
The coupling of superconductors with semiconductors has attracted significant interest recently, owing to the ensuing Andreev physics which, in combination with spin-orbit interaction and lifting of the spin degeneracy, can lead to non-trivial spin textures and could allow to explore exotic phases of matter.
Indeed, hybrid \SSm \ devices have become a prominent platform for engineering topological superconductivity, a key step towards fault-tolerant quantum computing~\cite{Kitaev2001,Nayak2008,Aguado:RNC17,Lutchyn:NRM18}.
In addition, such hybrid devices have been used to realize electrically controllable Josephson-junction qubits and they find application in the long-range coupling of spin qubits~\cite{Xiang2006,Lee:NN14,Ridderbos2019,Casparis2018,junger2020,larsen2015,larsen2020,luthi2018,Petersson2012,Burkard2020}.

Recent advancements in material science and fabrication have lead to a resurgence of interest in germanium~\cite{Xiang2006,Watzinger2018,Ridderbos2019,Ridderbos2020}. Indeed, hole gases in Ge offer several key physical properties such as inherent spin-orbit interaction, low hyperfine interaction and electrically tunable g-factors due to the carrier states originating from the valence band. In particular, the prospect of compatibility with existing Si foundry makes planar Ge a favourable platform for future quantum technologies~\cite{Scappucci2020}.
Recent breakthroughs with Ge-based spin qubits and hybrid \SSm \ devices underline its strong potential~\cite{HendrickxFour,Hendrickx2018,Hendrickx2019,Hendrickx2020,grenoble:ge}. 

For Ge-based \SSm \ devices, Al remains so far the foremost choice as a superconductor since it typically yields highly transparent contacts.
However, the limited magnetic-field resilience of Al acts as a deterrent for exploring exotic condensed matter phases. Among other common choices, Nb and NbTiN offer a higher superconducting gap and magnetic resilience, but forming high-quality interfaces with semiconductors is very challenging with these materials.

Here, we demonstrate induced superconductivity in Ge quantum wells (QWs), overcoming the main challenges of low-transparency interfaces and limited magnetic-field resilience.
The technique we employ is to use Al to form highly transparent and low-disorder interfaces with the QW, and then contact the thin Al layer directly by Nb, thereby increasing the superconducting gap of Al.
We determine the resulting effective gap by investigating signatures of multiple Andreev reflection in an SNS-junction that was fabricated in this way.
We further characterize the junction by studying its critical current as a function of temperature and magnetic field, and find that all our observations suggest that we have a long mean free path in the QW (exceeding the junction length) and highly transparent S-Sm interfaces.
We markedly see a higher critical magnetic field and \IcRn \ product in comparison to solely Al-based devices. Moreover, we demonstrate superconducting phase control over our junctions, which could allow to devise $\Phi_{0}$-junctions and explore low magnetic-field topological superconductivity~\cite{Pientka2017}.

\section{Results}

\subsection{Josephson Field Effect Transistors}
\begin{figure*}[t]
	\begin{center}
		\includegraphics[]{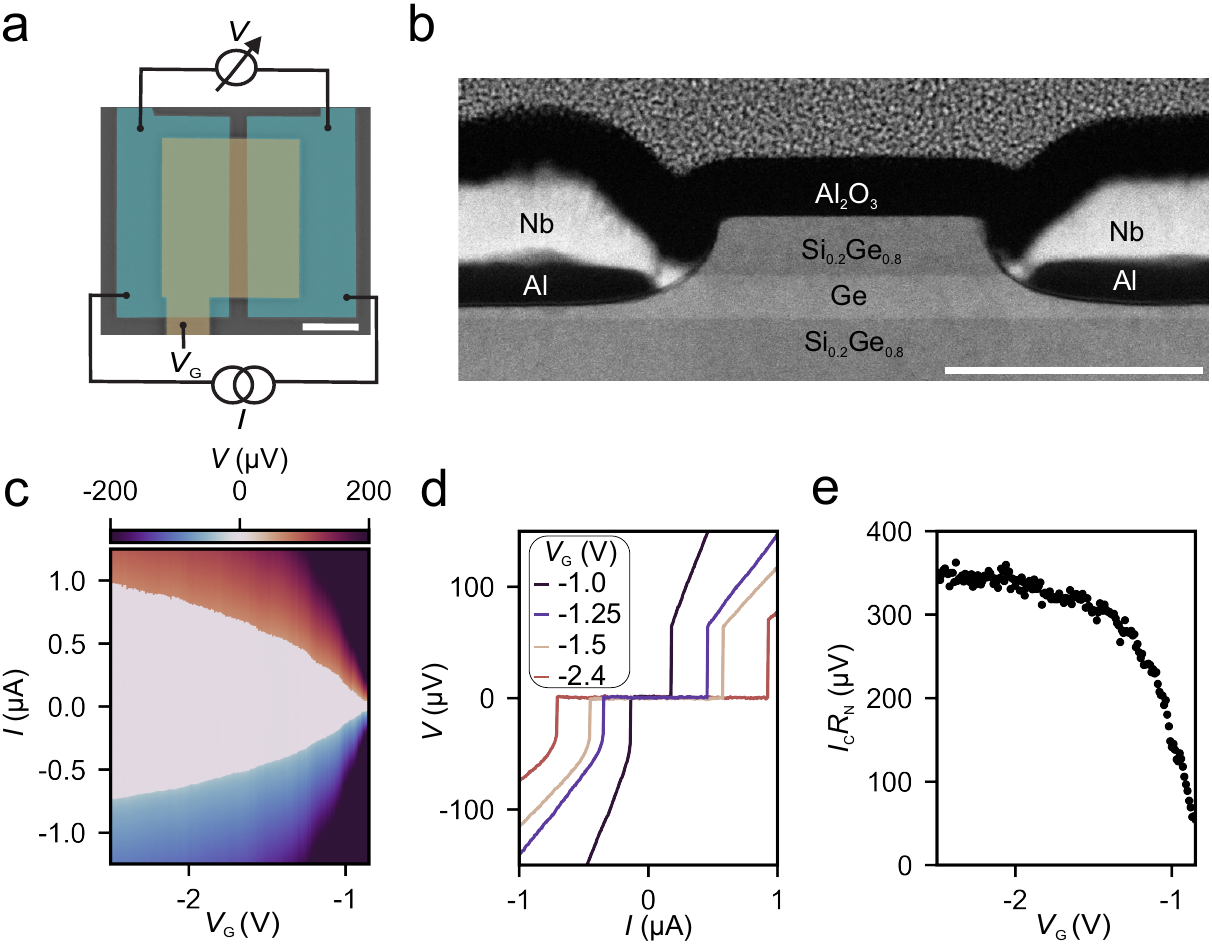}
		\hspace*{1cm}
		\caption{\textbf{(a)} False-colored SEM of a Ge JoFET with a top gate (yellow) accumulating a 2DHG between the two superconducting electrodes (blue). The scale bar is \SI{500}{nm}. \textbf{(b)} HAADF-STEM image of the cross-section of the JoFET with the Al layer directly contacting the Ge QW. The scale bar is \SI{100}{nm}. \textbf{(c)} $V$ measured across the JoFET versus $V_\textrm{G}$ and $I$. The device can be fully switched off at more positive voltages. \textbf{(d)} $V$ versus $I$ traces, extracted from \textbf{(c)}, highlighting the switching current at different $V_\textrm{G}$. \textbf{(e)} Dependence of the \IcRn \ product on $V_\textrm{G}$ as extracted from \textbf{(c)}.}
		\label{fig:Fig1}
	\end{center}
\end{figure*}

A Josephson Field Effect Transistor (JoFET), formed by sandwiching a semiconductor between two superconductors, allows to observe phase coherent Andreev transport reflecting the quality of the \SSm \ interface and the underlying transport in the semiconductor. We fabricate JoFETs with a strained Ge/SiGe heterostructure as a semiconducting weak-link. Densities of \SI{6d11}{cm^{-2}} and mobilities up to \SI{5d5}{cm^{2}/V.s}, leading to mean free paths \meanfreepath \ up to \SI{6}{\micro \meter}, are routinely achieved in nominally identical wafers~\cite{sammak:geqw}. \prettyref{fig:Fig1}a shows the false colored scanning electron microscope (SEM) image of a JoFET with the superconducting electrodes separated by a distance $L = $~\SI{150}{nm} and a top gate electrically isolated from the superconducting contacts by aluminium oxide. Further details on the fabrication of the devices can be found in the methods. \prettyref{fig:Fig1}b shows the cross-section High-Angle Annular Dark Field Scanning Transmission Electron Microscopy (HAADF-STEM) image of an identical JoFET where the Ge QW between two SiGe spacers is directly contacted by a thin film of Al to form a low-disorder and high transparency interface. Al itself is contacted by Nb resulting in a hybrid \SSSm \ junction. As observed in Electron Energy-Loss Spectroscopy (EELS) composition maps in Supp.~Fig.~4,  a region of \SI{3}{}--\SI{5}{nm} in the Nb layer is directly contacting the Ge QW. This Nb region is amorphous, as observed by atomic resolution HAADF-STEM, and partially oxidized, as shown by the EELS composition maps. The oxidation of Nb region is due to the influence of the Al$_\textrm{2}$O$_\textrm{3}$ layer grown on top of the device. A very thin halo of oxidation arises from the Al$_\textrm{2}$O$_\textrm{3}$, and extends through a few nanometers (\SI{3}{}--\SI{5}{nm}) in the Nb contact below forming the amorphous Nb oxidized region. The fact that Nb does not directly contact the Ge hole gas is further supported by the observation that devices made just with Nb superconducting electrodes did not show any current transport. Finally, we point out that the etching procedure produces a concave interface resulting in a larger segment of semiconducting weak-link than lithographically defined, potentially affecting the transport. A detailed overview of the HAADF-STEM and STEM-EELS analyses are presented in the Supplementary Material.  

The JoFETs are measured in a four-terminal current-biased configuration at a base temperature of \SI{20}{m K}. A top gate is used to tune the density of the underlying two-dimensional hole gas (2DHG) and we observe a gate-voltage-dependent switching current \Is \ of about \SI{1}{\micro A} at a negative gate voltage of \SI{-2.5}{V} (\prettyref{fig:Fig1}c and d). The clear dependence of \Is \ on the gate voltage provides evidence of Andreev transport occurring through the Ge QW. We expect the critical current \Ic \ to be almost equal to the experimentally measured \Is \ as the Josephson energy ${E}_\textrm{J} \approx \hbar\Is/2e \approx $~${k}_{\textrm{B}}$(\SI{2}-\SI{25}{K}) (${k}_\textrm{B}$ is the Boltzmann constant) is notably higher than the sample temperature, for the measured gate voltage range~\cite{Tinkham2004}. We further extract the characteristic \IcRn \ product, reaching up to \SI{360}{\micro V} as shown in \prettyref{fig:Fig1}e. In the Supp.~Fig.~5, we find \IcRn\ for the same fabrication process with Al as the sole superconductor reaching values up to \SI{50}{\micro V}, indicating a superior interface achieved between Ge and Al with our fabrication process compared to earlier works~\cite{delft:ge, grenoble:ge}. Harnessing the high quality \SSm \ interface, we enhance the superconducting properties of Al, and the hybrid devices, by contacting the Al layer directly with Nb~\cite{klapwijk:al-nb}. Therefore, we attribute the large \IcRn \ product to the combination of enhancement of the superconducting gap of Al due to contact with Nb and transparent Al-Ge interfaces.

\begin{figure}[t]
	\begin{center}
	\hspace*{-0.75cm}
		\includegraphics[]{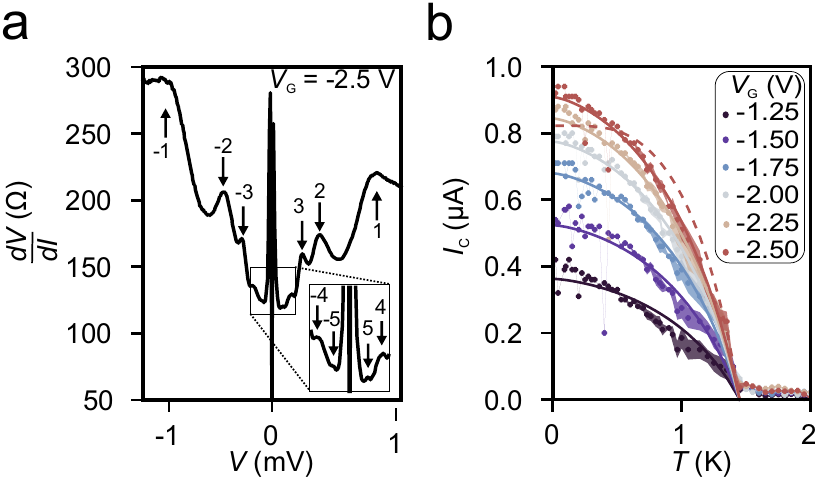}
		\caption{\textbf{(a)} Differential resistance $dV/dI$ versus voltage $V$ at $T=$~\SI{20}{\milli K} and $B=$~\SI{0}{T}, showing MAR peaks up to fifth order. The inset shows the higher order MAR features observed at lower values of $V$. 
		\textbf{(b)} Measured temperature dependence of the critical current \Ic\ for various top gate voltages \Vg\ (solid points, measurement error is indicated by the shaded regions). The lines present theoretical curves for a short junction~\cite{Beenakker1991a}; the solid lines assume a ballistic junction, the dashed line a diffusive junction (scaled to best match the data for $V_{\rm G}=$~\SI{-2.5}{V}.
        }
		\label{fig:Fig2}
	\end{center}
\end{figure}

\subsection{Multiple Andreev Reflection}

To characterize the JoFET in more detail, we measure its differential resistance $dV/dI$ versus the voltage $V$ over the junction.
The result is presented in \prettyref{fig:Fig2}a, showing a series of subgap features at finite voltages, indicated by the arrows.
We associate these features with the onset of Multiple Andreev Reflection (MAR) processes, which are expected to appear at voltages ${V} = {2}\Del/ne$, where $n$ is the number of times a quasiparticle is successively Andreev reflected and \Del \ is the superconducting gap of the electrodes~\cite{Klapwijk1982}.
In this way, we deduce \Del\ $\approx$~\SI{486}{\micro eV} (see Supp.~Fig.~6), which is markedly higher than that of bare Al ($\approx$ \SI{180}{\micro eV}) but lower than that of bare Nb ($\approx$~\SI{1.5}{meV}), providing further evidence of enlargement of the gap in the Al layer due to the proximity to Nb~\cite{Matthias1963}.
Using this value for \Del\ we find $e$\IcRn$/$\Del\ $\approx 0.75$, which is comparable to what was found for similar highly transparent S'-S-Sm-S-S' heterostructures made from NbTi, Al, and InAs~\cite{Drachmann2017,Kjaergaard2017}, but lower than the universal value of $\pi$ expected for short clean junctions.

The clear signatures of MAR suggest that the coherence length at low temperature $\xi_{\rm N}$ in the Ge QW is larger than $L$, and it also puts a lower bound on the inelastic scattering length of $l_\phi > 5L =$~\SI{750}{nm}.
Furthermore, the fact that the MAR features appear as \emph{peaks} in the resistance indicates that there is a high probability of Andreev reflection at the interfaces, i.e., that we have transparent \SSm\ contacts~\cite{Averin1995,Kjaergaard2017}.
This is not inconsistent with the magnitude of the excess current $I_{\rm ex} =$~\SI{2.3}{\micro A} (extracted in Supp.~Fig.~S8), which yields $e$\IexRn$/$\Del~$\approx 1.9$.
Using the Octavio-Blonder-Tinkham-Klapwijk model~\cite{flensberg1988, octavio1983} this would correspond to a barrier strength of $Z\approx 0.3$, translating to an average transparency of the junction of $\sim$~\SI{90}{\percent}.

\subsection{Temperature dependence}
\label{sec:tempdep}

We can obtain further information about the JoFET by investigating the temperature dependence of the critical current through the junction.
In \prettyref{fig:Fig2}b we plot $I_{\rm C}$ as a function of temperature for six different top gate voltages $V_{\rm G}$ (solid points).
One feature that stands out is that for all six traces the critical current drops to zero at the same temperature, which is approximately \SI{1.45}{K}.

We compare this temperature with the critical temperature one would expect in a simple BCS framework for the superconducting electrodes based on the measured gap, \Del$/1.76\,k_{\rm B} \approx$~\SI{3.2}{K}, and see that it is more than a factor 2 smaller.
In principle, this could indicate that $\xi_T$ (the length scale over which coherence is lost due to finite temperature) becomes smaller than $L$ already at intermediate $T$, before superconductivity in the electrodes is destroyed.
Indeed, in a junction that is not in the short-junction limit, i.e., when $L \gtrsim \xi_{\rm N}$, one expects an exponential suppression of the critical current when $L$ becomes larger than $\xi_T$,
manifesting itself as $I_{\rm C} \propto e^{-2\pi k_{\rm B} T L/\hbar v_{\rm F}}$ for a clean junction ($l_{\mathrm e} \gg L$, which is the limit we believe to be in, at least for the lowest top gate voltages) or $I_{\rm C} \propto e^{-\sqrt{2\pi k_{\rm B} T L^2/\frac{1}{2}\hbar v_{\rm F}l_{\mathrm e}}}$ for a dirty junction ($l_{\mathrm e} \ll L$)~\cite{kresin:thouless,Note1}.

However, two aspects of the data shown in \prettyref{fig:Fig2}b are inconsistent with this interpretation:
(i)~The vanishing of $I_{\rm C}$ at $T=$~\SI{1.45}{K} is too abrupt to fit either of the exponential functions very well.
(ii)~More importantly, the gate voltage $V_{\rm G}$ directly controls the hole density in the Ge QW~\cite{sammak:geqw} and thereby the Fermi velocity $v_{\rm F}$. This should result in a strong dependence on $V_{\rm G}$ of $\xi_T$ and thus the temperature where $I_{\rm C}$ becomes suppressed, which is clearly absent in the data.
We thus conclude that for all temperatures of interest we are most likely in the short-junction limit, and the vanishing of all supercurrent at \SI{1.45}{K} is due to the gap closing in the hybrid superconducting contacts.

We test whether the temperature-dependence of $I_{\rm C}(T)$ can qualitatively agree with the theory for short SNS-junctions ($L \ll \xi_{\rm N}$)~\cite{Beenakker1991,Beenakker1991a}.
Motivated by the long mean free path reported for our QW, we assume a clean junction ($L \ll l_{\mathrm e}$), for which the theory predicts
\begin{align}
I_{\rm C} \propto \max_\phi \left\{ \Del(T)\sin \frac{\phi}{2} \tanh\left( \frac{\Del(T)\cos\frac{\phi}{2}}{2 k_{\rm B}T} \right) \right\},
\label{eq:ic}
\end{align}
where ${\rm max}_\phi$ indicates maximization over the superconducting phase difference $\phi$.
Assuming for simplicity the BCS-like temperature dependence $\Del(T) = \Del(0) \tanh[ 1.74\sqrt{(T_c/T)-1} ]$, with $\Del(0) =$~\SI{486}{\micro eV} and $T_c =$~\SI{1.45}{K}, we scale Eq.~(\ref{eq:ic}) to fit the six traces in \prettyref{fig:Fig2}b; the result is plotted as solid lines.
Especially at more negative top gate voltages the curve given by Eq.~(\ref{eq:ic}) agrees well with the data.
To contrast this, we can also assume a short \emph{diffusive} junction ($l_{\mathrm e} \ll L \ll \xi_{\rm N}$) and average the general expression for the supercurrent given in Ref.~\cite{Beenakker1991a} over the so-called Dorokhov probability distribution for the transmission eigenvalues of a diffusive conductor~\cite{Dorokhov1984,Nazarov1994,Golubov2004}.
Extracting the critical current and scaling the resulting curve to best fit the data in that case yields the dashed line in \prettyref{fig:Fig2}b (for $V_{\rm G} =$~\SI{-2.5}{V}), which clearly agrees less well with our data.
We conclude that our temperature-dependent data, especially those at more negative gate voltages, are most consistent with the short and clean limit, where both $l_{\mathrm e}$ and $\xi_{\rm N}$ are larger than $L$.
We note that this is consistent with material properties reported for nominally identical Ge QWs~\cite{sammak:geqw}, where they found a mean free path up to \SI{6}{\micro m} and densities up to $6\times 10^{11}\,{\rm cm}^{-2}$ which (assuming $m^* = 0.1\,m_e$ and a strictly two-dimensional hole gas) yields $\hbar v_{\rm F}/\Del \approx$~\SI{300}{nm}.

\subsection{Magnetic field dependence}

\begin{figure}[t]
	\begin{center}
		\hspace*{-0.75cm}
		\includegraphics[]{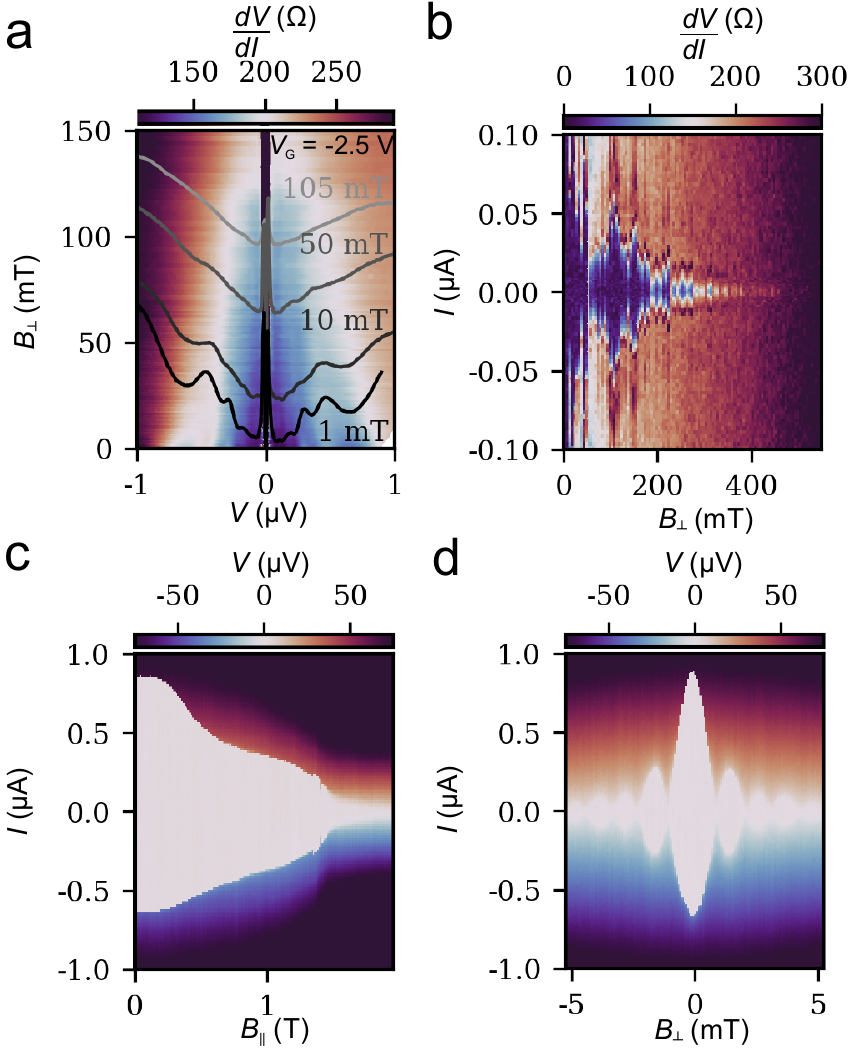}
		\caption{
			\textbf{(a)} Differential resistance $dV/dI$ versus perpendicular magnetic field $B_{\perp}$ and voltage $V$ measured over the junction. The overlying line traces show the evolution of the MAR features with $B_{\perp}$.
			\textbf{(b)} $dV/dI$ versus $B_{\perp}$ and bias current $I$, used to estimate the perpendicular critical magnetic field. The complex dependence of the switching current is due to the combination of the Fraunhofer effect and the drop of the switching current as the magnetic field is increasing.
			\textbf{(c)}~$V$ versus in-plane magnetic field $B_{\parallel}$ and $I$, used to estimate the in-plane critical field.
			\textbf{(d)} $V$ versus $B_{\perp}$ and $I$ at small applied magnetic fields, showing a Fraunhofer-like pattern.}
		\label{fig:Fig3}
	\end{center}
\end{figure}

We now turn our attention to the magnetic-field-dependent behavior of the JoFET.
In \prettyref{fig:Fig3}a we show its differential resistance $dV/dI$ versus the perpendicularly applied magnetic field $B_\perp$ and voltage $V$; we overlaid the data with four traces at the field strengths that are indicated in the plot.
We see that the MAR features that are clearly visible at low magnetic field (cf.~\prettyref{fig:Fig2}a) evolve to lower voltage with increasing magnetic field, indicating the decay of the superconducting gap due to the magnetic field~\cite{parks1968}.
In \prettyref{fig:Fig3}b we plot $dV/dI$ as a function of bias current $I$ and $B_\perp$, where we increase the magnetic field to higher values.
This allows us to find the critical perpendicular magnetic field for which the supercurrent vanishes, $B_{\perp, \textrm{C}} \approx$~\SI{460}{mT}.
The parallel critical magnetic field  $B_{\parallel, \textrm{C}} \approx$~\SI{1.8}{T} (as extracted from \prettyref{fig:Fig3}c) is almost four times higher, which is expected since the thickness of the superconducting electrodes is much smaller than their width.
The observed high magnetic-field resilience paves the way for exploring the interplay of magnetic effects in Ge with induced superconductivity and integration of disparate qubits such as spin qubits and gatemons on the same chip.

Finally, in \prettyref{fig:Fig3}d, we investigate the $I$-$V$ characteristics of the junction at small perpendicular fields, up to $\approx$~\SI{5}{\milli T}.
We find a clear Fraunhofer-like modulation of the switching current, confirming the coupling between the two superconducting leads through Andreev transport.
The observed symmetry of the pattern for positive and negative values of ${B}_{\perp}$ suggests low disorder in the Ge QW~\cite{rasmussen2016}, which is again consistent with our conclusions from the data shown in the previous Section.
We note that based on the lithographic dimensions of the junction, a magnetic field of \SI{6.9}{mT} should correspond to one magnetic flux quantum $h/2e$ threading through the junction area.
However, from the Fraunhofer pattern, we extract a magnetic field of \SI{0.8}{mT}, almost 9 times smaller than expected.
We attribute this difference to flux focusing of the applied magnetic field caused by the Meissner effect in the superconducting contacts~\cite{Khukhareva1963}.

\subsection{Multi-JoFET SQUID and CPR}
\begin{figure}[t]
	\begin{center}
		    \hspace*{-0.75cm}
		\includegraphics[]{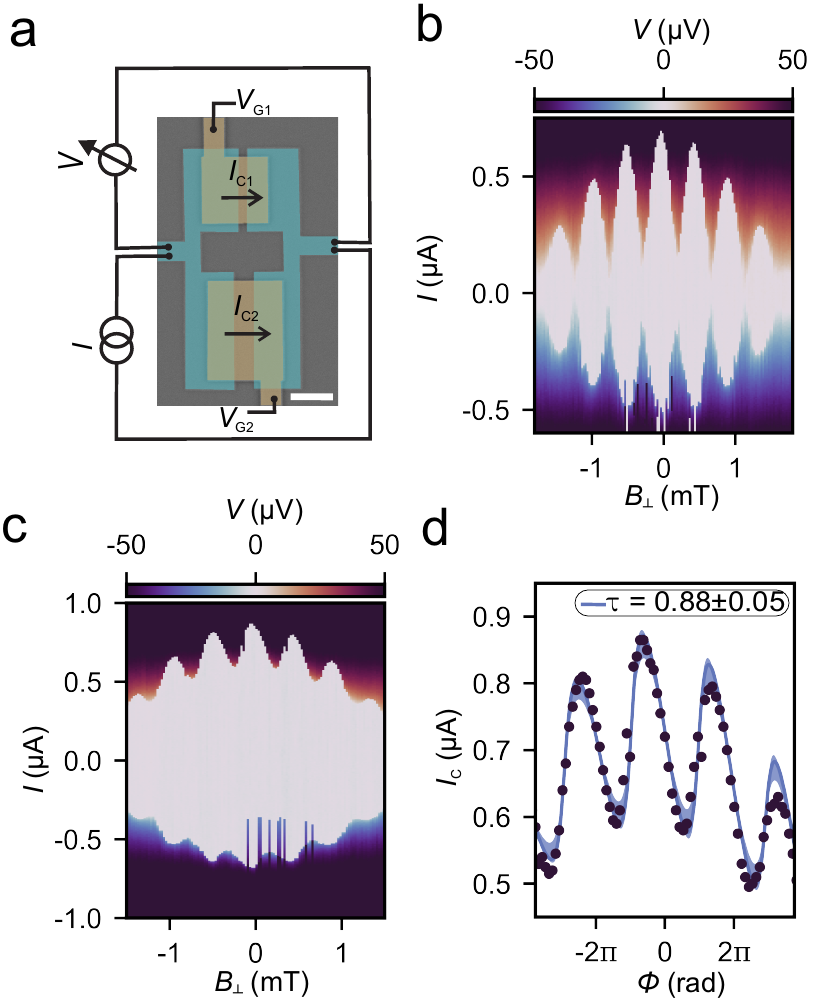}
		\caption{
			\textbf{(a)} False-colored SEM image of the two-JoFET asymmetric SQUID with channel lengths of \SI{150}{nm} and \SI{350}{nm}. The scale bar is \SI{1}{\micro m}.
			\textbf{(b)} $V_\textrm{G1} = $\SI{-1.245}{V} and $V_\textrm{G2} = $\SI{-9}{V} results in equal critical currents, yielding SQUID-like oscillations of the total critical current as a function of $B_\perp$.
			\textbf{(c)} $V_\textrm{G1}$ = \SI{-8}{V} and $V_\textrm{G2}$ = \SI{-1.4}{V} makes the superconducting phase drop mainly over JoFET2, allowing us to associate the oscillations in the critical current with the current-phase relationship of JoFET2.
			\textbf{(d)} Critical current extracted from \textbf{(c)} (solid points). The blue line shows a fit of the oscillations to the CPR given in Eq.~(\ref{eq:cpr}), yielding \texttau \ $ = 0.88 \pm 0.05$. The shaded region indicates the error in  \texttau.}
		\label{fig:Fig4}
	\end{center}
\end{figure}

Combining two JoFETs, we next explore interference patterns arising through the control over the superconducting phase difference using a perpendicular magnetic field.
\prettyref{fig:Fig4}a shows an asymmetric SQUID combining JoFETs with channel lengths \SI{150}{nm} (JoFET1) and \SI{350}{nm} (JoFET2).
The asymmetric channel lengths and individual top gate voltages allow tuning the double-JoFET device to various regimes, ranging from a conventional SQUID to a superconducting phase control device.

We investigate its operation in a four-probe configuration, by applying a current and measuring the voltage difference between the SQUID arms under the application of a perpendicular magnetic field.
(The behavior of each of the individual junctions in the asymmetric SQUID can be found in Supp.~Fig.~9.)
When the top gate voltages are tuned to achieve equal critical currents in the two junctions, $I_{\rm C1} = I_{\rm C2}$, we observe periodic oscillations of the critical current reflecting the underlying modulation of the superconducting phase due to a perpendicular magnetic field (see \prettyref{fig:Fig4}b).
The modulation period of the oscillations $\approx$~\SI{370}{\micro T} corresponds to one magnetic flux quantum through an area of \SI{5.5}{\micro m^2}, different from the lithographically defined area of the superconducting ring \SI{1.8}{\micro m^2}.
This difference we again attribute to significant flux focusing due to the Meissner effect and a difference in the net resultant area due to the finite penetration depth, as in the previous section.

Tuning to a large ratio of the two critical currents allows phase-biasing the individual JoFET with the lower critical current.
This gives direct access to its current-phase relationship (CPR), which can provide information about the underlying interfaces and physical phenomena at play~\cite{Nichele2020, Szombati2016, Mayer2020, Assouline2019}.
In \prettyref{fig:Fig4}c the system is tuned such that ${I}_\textrm{C1} \approx 9\, {I}_\textrm{C2}$; in this situation the change in the superconducting phase difference due to the magnetic field can be assumed to drop mainly over JoFET2.
We thus extract the critical current from \prettyref{fig:Fig4}c, plotted as solid points in \prettyref{fig:Fig4}d, and associate the oscillations we observe with the CPR of JoFET2.
The supercurrent in a transparent S-Sm-S junction can have a CPR that differs significantly from the sinusoidal CPR expected for superconductor-insulator-superconductor junctions~\cite{Golubov2004}.
The CPR underlying the data in \prettyref{fig:Fig4}d indeed seems to be skewed, so to obtain a rough estimate for the transparency of JoFET2 we fit the oscillations in $I_{\rm C}(\phi)$ using the same short-junction model as before~\cite{Beenakker1991a,Nichele2020,Mayer2020},
\begin{align}
I_{\rm S}(\phi) \propto \frac{\sin\phi}{\sqrt{1- \textrm{\texttau} \sin^{2}\frac{\phi}{2}}}, 
\label{eq:cpr}
\end{align}
assuming zero temperature and introducing the average transparency  \texttau \ of all channels in the junction as a fit parameter.
Taking into account the change of critical currents of the individual JoFETs due to the slower Fraunhofer modulation, this produces the fit presented by the solid blue line in \prettyref{fig:Fig4}d, yielding a high transparency of  \texttau \ $ = 0.88 \pm 0.05$, consistent with our earlier conclusions.

\section{Conclusion}

In summary, we present a characterization of a new type of hybrid SNS-junction, where a Ge-based two-dimensional hole gas is contacted by two superconducting Al leads that are in turn proximitized by an extra layer of Nb.
From a clear series of multiple-Andreev-reflection peaks observed in the differential conductance at low temperature and zero magnetic field, we extract a gap of \SI{486}{\micro eV} for the contacts, which is indeed enhanced considerably compared to bare Al.
The qualitative manifestation of the MAR features, the temperature dependence of the critical current, the magnitude of the excess current, the magnetic field-dependent behavior of the junction, and its detailed current-phase relationship all indicate that our junctions are in the short and clean limit ($L \lesssim \xi_{\rm N}, l_{\mathrm e}$) over a significant range of top-gate voltages and that the S-Sm interfaces connecting the Ge QW to the superconducting leads are highly transparent, which presents a considerable improvement for the hybrid planar Ge platform.
We finally demonstrate the ability to phase-bias individual JoFETs, which could allow to investigate different proposals for Majorana physics.
The system we developed thus establishes Ge as a viable platform for exploring exotic phases and as a hybrid qubit platform for bringing together spin and superconducting qubits on the same chip.

\begin{acknowledgements}

This research and related results were made possible with the support of the NOMIS Foundation. This research was supported by the Scientific Service Units of IST Austria through resources provided by the MIBA Machine Shop and the nanofabrication facility, the European Union’s Horizon 2020 research and innovation program under the Marie Sklodowska-Curie grant agreement \#844511 and the Grant Agreement \#862046. ICN2 acknowledge funding from Generalitat de Catalunya 2017 SGR 327. ICN2 is supported by the Severo Ochoa program from Spanish MINECO (Grant No. SEV-2017-0706) and is funded by the CERCA Programme / Generalitat de Catalunya. Part of the present work has been performed in the framework of Universitat  Autònoma de Barcelona Materials Science PhD program. The HAADF-STEM microscopy was conducted in the Laboratorio de Microscopias Avanzadas at Instituto de  Nanociencia de Aragon-Universidad de Zaragoza. Authors acknowledge the LMA-INA for offering  access to their instruments and expertise.  We acknowledge support from CSIC Research Platform on Quantum Technologies PTI-001. This project has received funding from the European Union’s Horizon 2020 research and innovation programme under grant agreement No 823717 – ESTEEM3. M.B. acknowledges support from SUR Generalitat de Catalunya and the EU Social Fund; project ref. 2020 FI 00103. GS and MV acknowledge support through a projectruimte grant associated with the Netherlands Organization of Scientific Research (NWO).
JD acknowledges support through FRIPRO-project 274853, which is funded by the Research Council of Norway.

\end{acknowledgements}

\textbf{DATA AVAILABILITY}
All transport data included in this work will be available on the IST Austria repository.

\appendix
\section{Methods}
The \SI{16}{nm} Ge QW heterostructure was grown by reduced chemical vapor deposition. Further details on the growth procedure can be found in Ref.~\cite{sammak:geqw}. The devices are fabricated using a \SI{100}{\kilo e V} ebeam lithography system. First, a reactive ion plasma etching step, based on  SF$_{6}$-O$_{2}$-CHF$_{3}$, is used to define mesa structures of $\approx$\SI{60}{nm} depth. This is followed by the deposition of the superconducting contacts. Before metal evaporation, the same plasma is used to etch $\approx$~\SI{35}{nm} of the heterostructure to ensure a direct contact between the superconductor and the Ge QW. Then we clean the exposed Ge QW with a 10~s BHF dip which is followed by a SF$_{6}$ plasma based passivation to reduce the contact resistance~\cite{7116476}. A \SI{15}{nm} thick layer of Al and a \SI{30}{nm} thick layer of Nb forming the superconducting contacts is deposited. A $\approx$~20nm thick layer of aluminium oxide is added at \SI{150}{\degreeCelsius} by plasma atomic layer deposition, followed by a top-gate consisting of \SI{3}{nm } Ti and \SI{97}{nm} Pd. 

We fabricated ten JoFETs, out of which two were not working due to leakage current through the gate oxide and the rest showed qualitatively similar supercurrents and \IcRn \ products.

\bibliography{references}

\begin{thebibliography}{50}%
\makeatletter
\providecommand \@ifxundefined [1]{%
 \@ifx{#1\undefined}
}%
\providecommand \@ifnum [1]{%
 \ifnum #1\expandafter \@firstoftwo
 \else \expandafter \@secondoftwo
 \fi
}%
\providecommand \@ifx [1]{%
 \ifx #1\expandafter \@firstoftwo
 \else \expandafter \@secondoftwo
 \fi
}%
\providecommand \natexlab [1]{#1}%
\providecommand \enquote  [1]{``#1''}%
\providecommand \bibnamefont  [1]{#1}%
\providecommand \bibfnamefont [1]{#1}%
\providecommand \citenamefont [1]{#1}%
\providecommand \href@noop [0]{\@secondoftwo}%
\providecommand \href [0]{\begingroup \@sanitize@url \@href}%
\providecommand \@href[1]{\@@startlink{#1}\@@href}%
\providecommand \@@href[1]{\endgroup#1\@@endlink}%
\providecommand \@sanitize@url [0]{\catcode `\\12\catcode `\$12\catcode
  `\&12\catcode `\#12\catcode `\^12\catcode `\_12\catcode `\%12\relax}%
\providecommand \@@startlink[1]{}%
\providecommand \@@endlink[0]{}%
\providecommand \url  [0]{\begingroup\@sanitize@url \@url }%
\providecommand \@url [1]{\endgroup\@href {#1}{\urlprefix }}%
\providecommand \urlprefix  [0]{URL }%
\providecommand \Eprint [0]{\href }%
\providecommand \doibase [0]{http://dx.doi.org/}%
\providecommand \selectlanguage [0]{\@gobble}%
\providecommand \bibinfo  [0]{\@secondoftwo}%
\providecommand \bibfield  [0]{\@secondoftwo}%
\providecommand \translation [1]{[#1]}%
\providecommand \BibitemOpen [0]{}%
\providecommand \bibitemStop [0]{}%
\providecommand \bibitemNoStop [0]{.\EOS\space}%
\providecommand \EOS [0]{\spacefactor3000\relax}%
\providecommand \BibitemShut  [1]{\csname bibitem#1\endcsname}%
\let\auto@bib@innerbib\@empty
\bibitem [{\citenamefont {Kitaev}(2001)}]{Kitaev2001}%
  \BibitemOpen
  \bibfield  {author} {\bibinfo {author} {\bibfnamefont {A.~Y.}\ \bibnamefont
  {Kitaev}},\ }\href {http://iopscience.iop.org/1063-7869/44/10S/S29/}
  {\bibfield  {journal} {\bibinfo  {journal} {Phys. Usp.}\ }\textbf {\bibinfo
  {volume} {44}},\ \bibinfo {pages} {131} (\bibinfo {year} {2001})}\BibitemShut
  {NoStop}%
\bibitem [{\citenamefont {Nayak}\ \emph {et~al.}(2008)\citenamefont {Nayak},
  \citenamefont {Simon}, \citenamefont {Stern}, \citenamefont {Freedman},\ and\
  \citenamefont {{Das Sarma}}}]{Nayak2008}%
  \BibitemOpen
  \bibfield  {author} {\bibinfo {author} {\bibfnamefont {C.}~\bibnamefont
  {Nayak}}, \bibinfo {author} {\bibfnamefont {S.~H.}\ \bibnamefont {Simon}},
  \bibinfo {author} {\bibfnamefont {A.}~\bibnamefont {Stern}}, \bibinfo
  {author} {\bibfnamefont {M.~H.}\ \bibnamefont {Freedman}}, \ and\ \bibinfo
  {author} {\bibfnamefont {S.}~\bibnamefont {{Das Sarma}}},\ }\href {\doibase
  10.1103/RevModPhys.80.1083} {\bibfield  {journal} {\bibinfo  {journal} {Rev.
  Mod. Phys.}\ }\textbf {\bibinfo {volume} {80}},\ \bibinfo {pages} {1083}
  (\bibinfo {year} {2008})}\BibitemShut {NoStop}%
\bibitem [{\citenamefont {Aguado}(2017)}]{Aguado:RNC17}%
  \BibitemOpen
  \bibfield  {author} {\bibinfo {author} {\bibfnamefont {R.}~\bibnamefont
  {Aguado}},\ }\href {\doibase 10.1393/ncr/i2017-10141-9} {\bibfield  {journal}
  {\bibinfo  {journal} {Riv. Nuovo Cimento}\ }\textbf {\bibinfo {volume}
  {40}},\ \bibinfo {pages} {523} (\bibinfo {year} {2017})}\BibitemShut
  {NoStop}%
\bibitem [{\citenamefont {Lutchyn}\ \emph {et~al.}(2018)\citenamefont
  {Lutchyn}, \citenamefont {Bakkers}, \citenamefont {Kouwenhoven},
  \citenamefont {Krogstrup}, \citenamefont {Marcus},\ and\ \citenamefont
  {Oreg}}]{Lutchyn:NRM18}%
  \BibitemOpen
  \bibfield  {author} {\bibinfo {author} {\bibfnamefont {R.~M.}\ \bibnamefont
  {Lutchyn}}, \bibinfo {author} {\bibfnamefont {E.~P. A.~M.}\ \bibnamefont
  {Bakkers}}, \bibinfo {author} {\bibfnamefont {L.~P.}\ \bibnamefont
  {Kouwenhoven}}, \bibinfo {author} {\bibfnamefont {P.}~\bibnamefont
  {Krogstrup}}, \bibinfo {author} {\bibfnamefont {C.~M.}\ \bibnamefont
  {Marcus}}, \ and\ \bibinfo {author} {\bibfnamefont {Y.}~\bibnamefont
  {Oreg}},\ }\href {\doibase 10.1038/s41578-018-0003-1} {\bibfield  {journal}
  {\bibinfo  {journal} {Nat. Rev. Mater.}\ }\textbf {\bibinfo {volume} {3}},\
  \bibinfo {pages} {52} (\bibinfo {year} {2018})}\BibitemShut {NoStop}%
\bibitem [{\citenamefont {Xiang}\ \emph {et~al.}(2006)\citenamefont {Xiang},
  \citenamefont {Vidan}, \citenamefont {Tinkham}, \citenamefont {Westervelt},\
  and\ \citenamefont {Lieber}}]{Xiang2006}%
  \BibitemOpen
  \bibfield  {author} {\bibinfo {author} {\bibfnamefont {J.}~\bibnamefont
  {Xiang}}, \bibinfo {author} {\bibfnamefont {A.}~\bibnamefont {Vidan}},
  \bibinfo {author} {\bibfnamefont {M.}~\bibnamefont {Tinkham}}, \bibinfo
  {author} {\bibfnamefont {R.~M.}\ \bibnamefont {Westervelt}}, \ and\ \bibinfo
  {author} {\bibfnamefont {C.~M.}\ \bibnamefont {Lieber}},\ }\href {\doibase
  10.1038/nnano.2006.140} {\bibfield  {journal} {\bibinfo  {journal} {Nature
  Nanotechnology}\ }\textbf {\bibinfo {volume} {1}},\ \bibinfo {pages} {208}
  (\bibinfo {year} {2006})}\BibitemShut {NoStop}%
\bibitem [{\citenamefont {Lee}\ \emph {et~al.}(2014)\citenamefont {Lee},
  \citenamefont {Jiang}, \citenamefont {Houzet}, \citenamefont {Aguado},
  \citenamefont {Lieber},\ and\ \citenamefont {De~Franceschi}}]{Lee:NN14}%
  \BibitemOpen
  \bibfield  {author} {\bibinfo {author} {\bibfnamefont {E.~J.~H.}\
  \bibnamefont {Lee}}, \bibinfo {author} {\bibfnamefont {X.}~\bibnamefont
  {Jiang}}, \bibinfo {author} {\bibfnamefont {M.}~\bibnamefont {Houzet}},
  \bibinfo {author} {\bibfnamefont {R.}~\bibnamefont {Aguado}}, \bibinfo
  {author} {\bibfnamefont {C.~M.}\ \bibnamefont {Lieber}}, \ and\ \bibinfo
  {author} {\bibfnamefont {S.}~\bibnamefont {De~Franceschi}},\ }\href {\doibase
  10.1038/nnano.2013.267} {\bibfield  {journal} {\bibinfo  {journal} {Nature
  Nanotechnology}\ }\textbf {\bibinfo {volume} {9}},\ \bibinfo {pages} {79}
  (\bibinfo {year} {2014})}\BibitemShut {NoStop}%
\bibitem [{\citenamefont {Ridderbos}\ \emph {et~al.}(2019)\citenamefont
  {Ridderbos}, \citenamefont {Brauns}, \citenamefont {Li}, \citenamefont
  {Bakkers}, \citenamefont {Brinkman}, \citenamefont {van~der Wiel},\ and\
  \citenamefont {Zwanenburg}}]{Ridderbos2019}%
  \BibitemOpen
  \bibfield  {author} {\bibinfo {author} {\bibfnamefont {J.}~\bibnamefont
  {Ridderbos}}, \bibinfo {author} {\bibfnamefont {M.}~\bibnamefont {Brauns}},
  \bibinfo {author} {\bibfnamefont {A.}~\bibnamefont {Li}}, \bibinfo {author}
  {\bibfnamefont {E.~P. A.~M.}\ \bibnamefont {Bakkers}}, \bibinfo {author}
  {\bibfnamefont {A.}~\bibnamefont {Brinkman}}, \bibinfo {author}
  {\bibfnamefont {W.~G.}\ \bibnamefont {van~der Wiel}}, \ and\ \bibinfo
  {author} {\bibfnamefont {F.~A.}\ \bibnamefont {Zwanenburg}},\ }\href
  {\doibase 10.1103/PhysRevMaterials.3.084803} {\bibfield  {journal} {\bibinfo
  {journal} {Phys. Rev. Materials}\ }\textbf {\bibinfo {volume} {3}},\ \bibinfo
  {pages} {084803} (\bibinfo {year} {2019})}\BibitemShut {NoStop}%
\bibitem [{\citenamefont {Casparis}\ \emph {et~al.}(2018)\citenamefont
  {Casparis}, \citenamefont {Connolly}, \citenamefont {Kjaergaard},
  \citenamefont {Pearson}, \citenamefont {Kringh{\o}j}, \citenamefont {Larsen},
  \citenamefont {Kuemmeth}, \citenamefont {Wang}, \citenamefont {Thomas},
  \citenamefont {Gronin}, \citenamefont {Gardner}, \citenamefont {Manfra},
  \citenamefont {Marcus},\ and\ \citenamefont {Petersson}}]{Casparis2018}%
  \BibitemOpen
  \bibfield  {author} {\bibinfo {author} {\bibfnamefont {L.}~\bibnamefont
  {Casparis}}, \bibinfo {author} {\bibfnamefont {M.~R.}\ \bibnamefont
  {Connolly}}, \bibinfo {author} {\bibfnamefont {M.}~\bibnamefont
  {Kjaergaard}}, \bibinfo {author} {\bibfnamefont {N.~J.}\ \bibnamefont
  {Pearson}}, \bibinfo {author} {\bibfnamefont {A.}~\bibnamefont
  {Kringh{\o}j}}, \bibinfo {author} {\bibfnamefont {T.~W.}\ \bibnamefont
  {Larsen}}, \bibinfo {author} {\bibfnamefont {F.}~\bibnamefont {Kuemmeth}},
  \bibinfo {author} {\bibfnamefont {T.}~\bibnamefont {Wang}}, \bibinfo {author}
  {\bibfnamefont {C.}~\bibnamefont {Thomas}}, \bibinfo {author} {\bibfnamefont
  {S.}~\bibnamefont {Gronin}}, \bibinfo {author} {\bibfnamefont {G.~C.}\
  \bibnamefont {Gardner}}, \bibinfo {author} {\bibfnamefont {M.~J.}\
  \bibnamefont {Manfra}}, \bibinfo {author} {\bibfnamefont {C.~M.}\
  \bibnamefont {Marcus}}, \ and\ \bibinfo {author} {\bibfnamefont {K.~D.}\
  \bibnamefont {Petersson}},\ }\href {\doibase 10.1038/s41565-018-0207-y}
  {\bibfield  {journal} {\bibinfo  {journal} {Nature Nanotechnology}\ }\textbf
  {\bibinfo {volume} {13}},\ \bibinfo {pages} {915} (\bibinfo {year}
  {2018})}\BibitemShut {NoStop}%
\bibitem [{\citenamefont {J\"unger}\ \emph {et~al.}(2020)\citenamefont
  {J\"unger}, \citenamefont {Delagrange}, \citenamefont {Chevallier},
  \citenamefont {Lehmann}, \citenamefont {Dick}, \citenamefont {Thelander},
  \citenamefont {Klinovaja}, \citenamefont {Loss}, \citenamefont
  {Baumgartner},\ and\ \citenamefont {Sch\"onenberger}}]{junger2020}%
  \BibitemOpen
  \bibfield  {author} {\bibinfo {author} {\bibfnamefont {C.}~\bibnamefont
  {J\"unger}}, \bibinfo {author} {\bibfnamefont {R.}~\bibnamefont
  {Delagrange}}, \bibinfo {author} {\bibfnamefont {D.}~\bibnamefont
  {Chevallier}}, \bibinfo {author} {\bibfnamefont {S.}~\bibnamefont {Lehmann}},
  \bibinfo {author} {\bibfnamefont {K.~A.}\ \bibnamefont {Dick}}, \bibinfo
  {author} {\bibfnamefont {C.}~\bibnamefont {Thelander}}, \bibinfo {author}
  {\bibfnamefont {J.}~\bibnamefont {Klinovaja}}, \bibinfo {author}
  {\bibfnamefont {D.}~\bibnamefont {Loss}}, \bibinfo {author} {\bibfnamefont
  {A.}~\bibnamefont {Baumgartner}}, \ and\ \bibinfo {author} {\bibfnamefont
  {C.}~\bibnamefont {Sch\"onenberger}},\ }\href
  {https://link.aps.org/doi/10.1103/PhysRevLett.125.017701} {\bibfield
  {journal} {\bibinfo  {journal} {Phys. Rev. Lett.}\ }\textbf {\bibinfo
  {volume} {125}},\ \bibinfo {pages} {017701} (\bibinfo {year}
  {2020})}\BibitemShut {NoStop}%
\bibitem [{\citenamefont {Larsen}\ \emph {et~al.}(2015)\citenamefont {Larsen},
  \citenamefont {Petersson}, \citenamefont {Kuemmeth}, \citenamefont
  {Jespersen}, \citenamefont {Krogstrup}, \citenamefont {Nyg\aa{}rd},\ and\
  \citenamefont {Marcus}}]{larsen2015}%
  \BibitemOpen
  \bibfield  {author} {\bibinfo {author} {\bibfnamefont {T.~W.}\ \bibnamefont
  {Larsen}}, \bibinfo {author} {\bibfnamefont {K.~D.}\ \bibnamefont
  {Petersson}}, \bibinfo {author} {\bibfnamefont {F.}~\bibnamefont {Kuemmeth}},
  \bibinfo {author} {\bibfnamefont {T.~S.}\ \bibnamefont {Jespersen}}, \bibinfo
  {author} {\bibfnamefont {P.}~\bibnamefont {Krogstrup}}, \bibinfo {author}
  {\bibfnamefont {J.}~\bibnamefont {Nyg\aa{}rd}}, \ and\ \bibinfo {author}
  {\bibfnamefont {C.~M.}\ \bibnamefont {Marcus}},\ }\href {\doibase
  10.1103/PhysRevLett.115.127001} {\bibfield  {journal} {\bibinfo  {journal}
  {Phys. Rev. Lett.}\ }\textbf {\bibinfo {volume} {115}},\ \bibinfo {pages}
  {127001} (\bibinfo {year} {2015})}\BibitemShut {NoStop}%
\bibitem [{\citenamefont {Larsen}\ \emph {et~al.}(2020)\citenamefont {Larsen},
  \citenamefont {Gershenson}, \citenamefont {Casparis}, \citenamefont
  {Kringh\o{}j}, \citenamefont {Pearson}, \citenamefont {McNeil}, \citenamefont
  {Kuemmeth}, \citenamefont {Krogstrup}, \citenamefont {Petersson},\ and\
  \citenamefont {Marcus}}]{larsen2020}%
  \BibitemOpen
  \bibfield  {author} {\bibinfo {author} {\bibfnamefont {T.~W.}\ \bibnamefont
  {Larsen}}, \bibinfo {author} {\bibfnamefont {M.~E.}\ \bibnamefont
  {Gershenson}}, \bibinfo {author} {\bibfnamefont {L.}~\bibnamefont
  {Casparis}}, \bibinfo {author} {\bibfnamefont {A.}~\bibnamefont
  {Kringh\o{}j}}, \bibinfo {author} {\bibfnamefont {N.~J.}\ \bibnamefont
  {Pearson}}, \bibinfo {author} {\bibfnamefont {R.~P.~G.}\ \bibnamefont
  {McNeil}}, \bibinfo {author} {\bibfnamefont {F.}~\bibnamefont {Kuemmeth}},
  \bibinfo {author} {\bibfnamefont {P.}~\bibnamefont {Krogstrup}}, \bibinfo
  {author} {\bibfnamefont {K.~D.}\ \bibnamefont {Petersson}}, \ and\ \bibinfo
  {author} {\bibfnamefont {C.~M.}\ \bibnamefont {Marcus}},\ }\href {\doibase
  10.1103/PhysRevLett.125.056801} {\bibfield  {journal} {\bibinfo  {journal}
  {Phys. Rev. Lett.}\ }\textbf {\bibinfo {volume} {125}},\ \bibinfo {pages}
  {056801} (\bibinfo {year} {2020})}\BibitemShut {NoStop}%
\bibitem [{\citenamefont {Luthi}\ \emph {et~al.}(2018)\citenamefont {Luthi},
  \citenamefont {Stavenga}, \citenamefont {Enzing}, \citenamefont {Bruno},
  \citenamefont {Dickel}, \citenamefont {Langford}, \citenamefont {Rol},
  \citenamefont {Jespersen}, \citenamefont {Nyg\aa{}rd}, \citenamefont
  {Krogstrup},\ and\ \citenamefont {DiCarlo}}]{luthi2018}%
  \BibitemOpen
  \bibfield  {author} {\bibinfo {author} {\bibfnamefont {F.}~\bibnamefont
  {Luthi}}, \bibinfo {author} {\bibfnamefont {T.}~\bibnamefont {Stavenga}},
  \bibinfo {author} {\bibfnamefont {O.~W.}\ \bibnamefont {Enzing}}, \bibinfo
  {author} {\bibfnamefont {A.}~\bibnamefont {Bruno}}, \bibinfo {author}
  {\bibfnamefont {C.}~\bibnamefont {Dickel}}, \bibinfo {author} {\bibfnamefont
  {N.~K.}\ \bibnamefont {Langford}}, \bibinfo {author} {\bibfnamefont {M.~A.}\
  \bibnamefont {Rol}}, \bibinfo {author} {\bibfnamefont {T.~S.}\ \bibnamefont
  {Jespersen}}, \bibinfo {author} {\bibfnamefont {J.}~\bibnamefont
  {Nyg\aa{}rd}}, \bibinfo {author} {\bibfnamefont {P.}~\bibnamefont
  {Krogstrup}}, \ and\ \bibinfo {author} {\bibfnamefont {L.}~\bibnamefont
  {DiCarlo}},\ }\href {\doibase 10.1103/PhysRevLett.120.100502} {\bibfield
  {journal} {\bibinfo  {journal} {Phys. Rev. Lett.}\ }\textbf {\bibinfo
  {volume} {120}},\ \bibinfo {pages} {100502} (\bibinfo {year}
  {2018})}\BibitemShut {NoStop}%
\bibitem [{\citenamefont {Petersson}\ \emph {et~al.}(2012)\citenamefont
  {Petersson}, \citenamefont {McFaul}, \citenamefont {Schroer}, \citenamefont
  {Jung}, \citenamefont {Taylor}, \citenamefont {Houck},\ and\ \citenamefont
  {Petta}}]{Petersson2012}%
  \BibitemOpen
  \bibfield  {author} {\bibinfo {author} {\bibfnamefont {K.~D.}\ \bibnamefont
  {Petersson}}, \bibinfo {author} {\bibfnamefont {L.~W.}\ \bibnamefont
  {McFaul}}, \bibinfo {author} {\bibfnamefont {M.~D.}\ \bibnamefont {Schroer}},
  \bibinfo {author} {\bibfnamefont {M.}~\bibnamefont {Jung}}, \bibinfo {author}
  {\bibfnamefont {J.~M.}\ \bibnamefont {Taylor}}, \bibinfo {author}
  {\bibfnamefont {A.~A.}\ \bibnamefont {Houck}}, \ and\ \bibinfo {author}
  {\bibfnamefont {J.~R.}\ \bibnamefont {Petta}},\ }\href {\doibase
  10.1038/nature11559} {\bibfield  {journal} {\bibinfo  {journal} {Nature}\
  }\textbf {\bibinfo {volume} {490}},\ \bibinfo {pages} {380} (\bibinfo {year}
  {2012})}\BibitemShut {NoStop}%
\bibitem [{\citenamefont {Burkard}\ \emph {et~al.}(2020)\citenamefont
  {Burkard}, \citenamefont {Gullans}, \citenamefont {Mi},\ and\ \citenamefont
  {Petta}}]{Burkard2020}%
  \BibitemOpen
  \bibfield  {author} {\bibinfo {author} {\bibfnamefont {G.}~\bibnamefont
  {Burkard}}, \bibinfo {author} {\bibfnamefont {M.~J.}\ \bibnamefont
  {Gullans}}, \bibinfo {author} {\bibfnamefont {X.}~\bibnamefont {Mi}}, \ and\
  \bibinfo {author} {\bibfnamefont {J.~R.}\ \bibnamefont {Petta}},\ }\href
  {\doibase 10.1038/s42254-019-0135-2} {\bibfield  {journal} {\bibinfo
  {journal} {Nature Reviews Physics}\ }\textbf {\bibinfo {volume} {2}},\
  \bibinfo {pages} {129} (\bibinfo {year} {2020})}\BibitemShut {NoStop}%
\bibitem [{\citenamefont {Watzinger}\ \emph {et~al.}(2018)\citenamefont
  {Watzinger}, \citenamefont {Kuku{\v{c}}ka}, \citenamefont
  {Vuku{\v{s}}i{\'{c}}}, \citenamefont {Gao}, \citenamefont {Wang},
  \citenamefont {Sch\"{a}ffler}, \citenamefont {Zhang},\ and\ \citenamefont
  {Katsaros}}]{Watzinger2018}%
  \BibitemOpen
  \bibfield  {author} {\bibinfo {author} {\bibfnamefont {H.}~\bibnamefont
  {Watzinger}}, \bibinfo {author} {\bibfnamefont {J.}~\bibnamefont
  {Kuku{\v{c}}ka}}, \bibinfo {author} {\bibfnamefont {L.}~\bibnamefont
  {Vuku{\v{s}}i{\'{c}}}}, \bibinfo {author} {\bibfnamefont {F.}~\bibnamefont
  {Gao}}, \bibinfo {author} {\bibfnamefont {T.}~\bibnamefont {Wang}}, \bibinfo
  {author} {\bibfnamefont {F.}~\bibnamefont {Sch\"{a}ffler}}, \bibinfo {author}
  {\bibfnamefont {J.-J.}\ \bibnamefont {Zhang}}, \ and\ \bibinfo {author}
  {\bibfnamefont {G.}~\bibnamefont {Katsaros}},\ }\href
  {https://doi.org/10.1038/s41467-018-06418-4} {\bibfield  {journal} {\bibinfo
  {journal} {Nature Communications}\ }\textbf {\bibinfo {volume} {9}},\
  \bibinfo {pages} {3902} (\bibinfo {year} {2018})}\BibitemShut {NoStop}%
\bibitem [{\citenamefont {Ridderbos}\ \emph {et~al.}(2020)\citenamefont
  {Ridderbos}, \citenamefont {Brauns}, \citenamefont {de~Vries}, \citenamefont
  {Shen}, \citenamefont {Li}, \citenamefont {K{\"o}lling}, \citenamefont
  {Verheijen}, \citenamefont {Brinkman}, \citenamefont {van~der Wiel},
  \citenamefont {Bakkers},\ and\ \citenamefont {Zwanenburg}}]{Ridderbos2020}%
  \BibitemOpen
  \bibfield  {author} {\bibinfo {author} {\bibfnamefont {J.}~\bibnamefont
  {Ridderbos}}, \bibinfo {author} {\bibfnamefont {M.}~\bibnamefont {Brauns}},
  \bibinfo {author} {\bibfnamefont {F.~K.}\ \bibnamefont {de~Vries}}, \bibinfo
  {author} {\bibfnamefont {J.}~\bibnamefont {Shen}}, \bibinfo {author}
  {\bibfnamefont {A.}~\bibnamefont {Li}}, \bibinfo {author} {\bibfnamefont
  {S.}~\bibnamefont {K{\"o}lling}}, \bibinfo {author} {\bibfnamefont {M.~A.}\
  \bibnamefont {Verheijen}}, \bibinfo {author} {\bibfnamefont {A.}~\bibnamefont
  {Brinkman}}, \bibinfo {author} {\bibfnamefont {W.~G.}\ \bibnamefont {van~der
  Wiel}}, \bibinfo {author} {\bibfnamefont {E.~P. A.~M.}\ \bibnamefont
  {Bakkers}}, \ and\ \bibinfo {author} {\bibfnamefont {F.~A.}\ \bibnamefont
  {Zwanenburg}},\ }\href {\doibase 10.1021/acs.nanolett.9b03438} {\bibfield
  {journal} {\bibinfo  {journal} {Nano Letters}\ }\textbf {\bibinfo {volume}
  {20}},\ \bibinfo {pages} {122} (\bibinfo {year} {2020})}\BibitemShut
  {NoStop}%
\bibitem [{\citenamefont {Scappucci}\ \emph {et~al.}(2020)\citenamefont
  {Scappucci}, \citenamefont {Kloeffel}, \citenamefont {Zwanenburg},
  \citenamefont {Loss}, \citenamefont {Myronov}, \citenamefont {Zhang},
  \citenamefont {De~Franceschi}, \citenamefont {Katsaros},\ and\ \citenamefont
  {Veldhorst}}]{Scappucci2020}%
  \BibitemOpen
  \bibfield  {author} {\bibinfo {author} {\bibfnamefont {G.}~\bibnamefont
  {Scappucci}}, \bibinfo {author} {\bibfnamefont {C.}~\bibnamefont {Kloeffel}},
  \bibinfo {author} {\bibfnamefont {F.~A.}\ \bibnamefont {Zwanenburg}},
  \bibinfo {author} {\bibfnamefont {D.}~\bibnamefont {Loss}}, \bibinfo {author}
  {\bibfnamefont {M.}~\bibnamefont {Myronov}}, \bibinfo {author} {\bibfnamefont
  {J.-J.}\ \bibnamefont {Zhang}}, \bibinfo {author} {\bibfnamefont
  {S.}~\bibnamefont {De~Franceschi}}, \bibinfo {author} {\bibfnamefont
  {G.}~\bibnamefont {Katsaros}}, \ and\ \bibinfo {author} {\bibfnamefont
  {M.}~\bibnamefont {Veldhorst}},\ }\href
  {https://doi.org/10.1038/s41578-020-00262-z} {\bibfield  {journal} {\bibinfo
  {journal} {Nature Reviews Materials}\ } (\bibinfo {year} {2020})}\BibitemShut
  {NoStop}%
\bibitem [{\citenamefont {{Hendrickx}}\ \emph {et~al.}(2020)\citenamefont
  {{Hendrickx}}, \citenamefont {{Lawrie}}, \citenamefont {{Russ}},
  \citenamefont {{van Riggelen}}, \citenamefont {{de Snoo}}, \citenamefont
  {{Schouten}}, \citenamefont {{Sammak}}, \citenamefont {{Scappucci}},\ and\
  \citenamefont {{Veldhorst}}}]{HendrickxFour}%
  \BibitemOpen
  \bibfield  {author} {\bibinfo {author} {\bibfnamefont {N.~W.}\ \bibnamefont
  {{Hendrickx}}}, \bibinfo {author} {\bibfnamefont {W.~I.~L.}\ \bibnamefont
  {{Lawrie}}}, \bibinfo {author} {\bibfnamefont {M.}~\bibnamefont {{Russ}}},
  \bibinfo {author} {\bibfnamefont {F.}~\bibnamefont {{van Riggelen}}},
  \bibinfo {author} {\bibfnamefont {S.~L.}\ \bibnamefont {{de Snoo}}}, \bibinfo
  {author} {\bibfnamefont {R.~N.}\ \bibnamefont {{Schouten}}}, \bibinfo
  {author} {\bibfnamefont {A.}~\bibnamefont {{Sammak}}}, \bibinfo {author}
  {\bibfnamefont {G.}~\bibnamefont {{Scappucci}}}, \ and\ \bibinfo {author}
  {\bibfnamefont {M.}~\bibnamefont {{Veldhorst}}},\ }\href
  {https://ui.adsabs.harvard.edu/abs/2020arXiv200904268H} {\ ,\ \bibinfo
  {pages} {arXiv:2009.04268} (\bibinfo {year} {2020})}\BibitemShut {NoStop}%
\bibitem [{\citenamefont {Hendrickx}\ \emph {et~al.}(2018)\citenamefont
  {Hendrickx}, \citenamefont {Franke}, \citenamefont {Sammak}, \citenamefont
  {Kouwenhoven}, \citenamefont {Sabbagh}, \citenamefont {Yeoh}, \citenamefont
  {Li}, \citenamefont {Tagliaferri}, \citenamefont {Virgilio}, \citenamefont
  {Capellini}, \citenamefont {Scappucci},\ and\ \citenamefont
  {Veldhorst}}]{Hendrickx2018}%
  \BibitemOpen
  \bibfield  {author} {\bibinfo {author} {\bibfnamefont {N.~W.}\ \bibnamefont
  {Hendrickx}}, \bibinfo {author} {\bibfnamefont {D.~P.}\ \bibnamefont
  {Franke}}, \bibinfo {author} {\bibfnamefont {A.}~\bibnamefont {Sammak}},
  \bibinfo {author} {\bibfnamefont {M.}~\bibnamefont {Kouwenhoven}}, \bibinfo
  {author} {\bibfnamefont {D.}~\bibnamefont {Sabbagh}}, \bibinfo {author}
  {\bibfnamefont {L.}~\bibnamefont {Yeoh}}, \bibinfo {author} {\bibfnamefont
  {R.}~\bibnamefont {Li}}, \bibinfo {author} {\bibfnamefont {M.~L.~V.}\
  \bibnamefont {Tagliaferri}}, \bibinfo {author} {\bibfnamefont
  {M.}~\bibnamefont {Virgilio}}, \bibinfo {author} {\bibfnamefont
  {G.}~\bibnamefont {Capellini}}, \bibinfo {author} {\bibfnamefont
  {G.}~\bibnamefont {Scappucci}}, \ and\ \bibinfo {author} {\bibfnamefont
  {M.}~\bibnamefont {Veldhorst}},\ }\href
  {https://doi.org/10.1038/s41467-018-05299-x} {\bibfield  {journal} {\bibinfo
  {journal} {Nature Communications}\ }\textbf {\bibinfo {volume} {9}},\
  \bibinfo {pages} {2835} (\bibinfo {year} {2018})}\BibitemShut {NoStop}%
\bibitem [{\citenamefont {Hendrickx}\ \emph
  {et~al.}(2019{\natexlab{a}})\citenamefont {Hendrickx}, \citenamefont
  {Tagliaferri}, \citenamefont {Kouwenhoven}, \citenamefont {Li}, \citenamefont
  {Franke}, \citenamefont {Sammak}, \citenamefont {Brinkman}, \citenamefont
  {Scappucci},\ and\ \citenamefont {Veldhorst}}]{Hendrickx2019}%
  \BibitemOpen
  \bibfield  {author} {\bibinfo {author} {\bibfnamefont {N.~W.}\ \bibnamefont
  {Hendrickx}}, \bibinfo {author} {\bibfnamefont {M.~L.~V.}\ \bibnamefont
  {Tagliaferri}}, \bibinfo {author} {\bibfnamefont {M.}~\bibnamefont
  {Kouwenhoven}}, \bibinfo {author} {\bibfnamefont {R.}~\bibnamefont {Li}},
  \bibinfo {author} {\bibfnamefont {D.~P.}\ \bibnamefont {Franke}}, \bibinfo
  {author} {\bibfnamefont {A.}~\bibnamefont {Sammak}}, \bibinfo {author}
  {\bibfnamefont {A.}~\bibnamefont {Brinkman}}, \bibinfo {author}
  {\bibfnamefont {G.}~\bibnamefont {Scappucci}}, \ and\ \bibinfo {author}
  {\bibfnamefont {M.}~\bibnamefont {Veldhorst}},\ }\href {\doibase
  10.1103/PhysRevB.99.075435} {\bibfield  {journal} {\bibinfo  {journal} {Phys.
  Rev. B}\ }\textbf {\bibinfo {volume} {99}},\ \bibinfo {pages} {075435}
  (\bibinfo {year} {2019}{\natexlab{a}})}\BibitemShut {NoStop}%
\bibitem [{\citenamefont {Hendrickx}\ \emph {et~al.}(2020)\citenamefont
  {Hendrickx}, \citenamefont {Franke}, \citenamefont {Sammak}, \citenamefont
  {Scappucci},\ and\ \citenamefont {Veldhorst}}]{Hendrickx2020}%
  \BibitemOpen
  \bibfield  {author} {\bibinfo {author} {\bibfnamefont {N.~W.}\ \bibnamefont
  {Hendrickx}}, \bibinfo {author} {\bibfnamefont {D.~P.}\ \bibnamefont
  {Franke}}, \bibinfo {author} {\bibfnamefont {A.}~\bibnamefont {Sammak}},
  \bibinfo {author} {\bibfnamefont {G.}~\bibnamefont {Scappucci}}, \ and\
  \bibinfo {author} {\bibfnamefont {M.}~\bibnamefont {Veldhorst}},\ }\href
  {\doibase 10.1038/s41586-019-1919-3} {\bibfield  {journal} {\bibinfo
  {journal} {Nature}\ }\textbf {\bibinfo {volume} {577}},\ \bibinfo {pages}
  {487} (\bibinfo {year} {2020})}\BibitemShut {NoStop}%
\bibitem [{\citenamefont {Vigneau}\ \emph {et~al.}(2019)\citenamefont
  {Vigneau}, \citenamefont {Mizokuchi}, \citenamefont {Zanuz}, \citenamefont
  {Huang}, \citenamefont {Tan}, \citenamefont {Maurand}, \citenamefont
  {Frolov}, \citenamefont {Sammak}, \citenamefont {Scappucci}, \citenamefont
  {Lefloch},\ and\ \citenamefont {De~Franceschi}}]{grenoble:ge}%
  \BibitemOpen
  \bibfield  {author} {\bibinfo {author} {\bibfnamefont {F.}~\bibnamefont
  {Vigneau}}, \bibinfo {author} {\bibfnamefont {R.}~\bibnamefont {Mizokuchi}},
  \bibinfo {author} {\bibfnamefont {D.~C.}\ \bibnamefont {Zanuz}}, \bibinfo
  {author} {\bibfnamefont {X.}~\bibnamefont {Huang}}, \bibinfo {author}
  {\bibfnamefont {S.}~\bibnamefont {Tan}}, \bibinfo {author} {\bibfnamefont
  {R.}~\bibnamefont {Maurand}}, \bibinfo {author} {\bibfnamefont
  {S.}~\bibnamefont {Frolov}}, \bibinfo {author} {\bibfnamefont
  {A.}~\bibnamefont {Sammak}}, \bibinfo {author} {\bibfnamefont
  {G.}~\bibnamefont {Scappucci}}, \bibinfo {author} {\bibfnamefont
  {F.}~\bibnamefont {Lefloch}}, \ and\ \bibinfo {author} {\bibfnamefont
  {S.}~\bibnamefont {De~Franceschi}},\ }\href
  {https://doi.org/10.1021/acs.nanolett.8b04275} {\bibfield  {journal}
  {\bibinfo  {journal} {Nano Letters}\ }\textbf {\bibinfo {volume} {19}},\
  \bibinfo {pages} {1023} (\bibinfo {year} {2019})}\BibitemShut {NoStop}%
\bibitem [{\citenamefont {Pientka}\ \emph {et~al.}(2017)\citenamefont
  {Pientka}, \citenamefont {Keselman}, \citenamefont {Berg}, \citenamefont
  {Yacoby}, \citenamefont {Stern},\ and\ \citenamefont
  {Halperin}}]{Pientka2017}%
  \BibitemOpen
  \bibfield  {author} {\bibinfo {author} {\bibfnamefont {F.}~\bibnamefont
  {Pientka}}, \bibinfo {author} {\bibfnamefont {A.}~\bibnamefont {Keselman}},
  \bibinfo {author} {\bibfnamefont {E.}~\bibnamefont {Berg}}, \bibinfo {author}
  {\bibfnamefont {A.}~\bibnamefont {Yacoby}}, \bibinfo {author} {\bibfnamefont
  {A.}~\bibnamefont {Stern}}, \ and\ \bibinfo {author} {\bibfnamefont {B.~I.}\
  \bibnamefont {Halperin}},\ }\href {\doibase 10.1103/PhysRevX.7.021032}
  {\bibfield  {journal} {\bibinfo  {journal} {Phys. Rev. X}\ }\textbf {\bibinfo
  {volume} {7}},\ \bibinfo {pages} {021032} (\bibinfo {year}
  {2017})}\BibitemShut {NoStop}%
\bibitem [{\citenamefont {Sammak}\ \emph {et~al.}(2019)\citenamefont {Sammak},
  \citenamefont {Sabbagh}, \citenamefont {Hendrickx}, \citenamefont {Lodari},
  \citenamefont {Paquelet~Wuetz}, \citenamefont {Tosato}, \citenamefont {Yeoh},
  \citenamefont {Bollani}, \citenamefont {Virgilio}, \citenamefont {Schubert},
  \citenamefont {Zaumseil}, \citenamefont {Capellini}, \citenamefont
  {Veldhorst},\ and\ \citenamefont {Scappucci}}]{sammak:geqw}%
  \BibitemOpen
  \bibfield  {author} {\bibinfo {author} {\bibfnamefont {A.}~\bibnamefont
  {Sammak}}, \bibinfo {author} {\bibfnamefont {D.}~\bibnamefont {Sabbagh}},
  \bibinfo {author} {\bibfnamefont {N.~W.}\ \bibnamefont {Hendrickx}}, \bibinfo
  {author} {\bibfnamefont {M.}~\bibnamefont {Lodari}}, \bibinfo {author}
  {\bibfnamefont {B.}~\bibnamefont {Paquelet~Wuetz}}, \bibinfo {author}
  {\bibfnamefont {A.}~\bibnamefont {Tosato}}, \bibinfo {author} {\bibfnamefont
  {L.}~\bibnamefont {Yeoh}}, \bibinfo {author} {\bibfnamefont {M.}~\bibnamefont
  {Bollani}}, \bibinfo {author} {\bibfnamefont {M.}~\bibnamefont {Virgilio}},
  \bibinfo {author} {\bibfnamefont {M.~A.}\ \bibnamefont {Schubert}}, \bibinfo
  {author} {\bibfnamefont {P.}~\bibnamefont {Zaumseil}}, \bibinfo {author}
  {\bibfnamefont {G.}~\bibnamefont {Capellini}}, \bibinfo {author}
  {\bibfnamefont {M.}~\bibnamefont {Veldhorst}}, \ and\ \bibinfo {author}
  {\bibfnamefont {G.}~\bibnamefont {Scappucci}},\ }\href {\doibase
  10.1002/adfm.201807613} {\bibfield  {journal} {\bibinfo  {journal} {Advanced
  Functional Materials}\ }\textbf {\bibinfo {volume} {29}},\ \bibinfo {pages}
  {1807613} (\bibinfo {year} {2019})}\BibitemShut {NoStop}%
\bibitem [{\citenamefont {Tinkham}(2004)}]{Tinkham2004}%
  \BibitemOpen
  \bibfield  {author} {\bibinfo {author} {\bibfnamefont {M.}~\bibnamefont
  {Tinkham}},\ }\href@noop {} {\emph {\bibinfo {title} {Introduction to
  Superconductivity -}}}\ (\bibinfo  {publisher} {Courier Corporation},\
  \bibinfo {address} {New York},\ \bibinfo {year} {2004})\BibitemShut {NoStop}%
\bibitem [{\citenamefont {Hendrickx}\ \emph
  {et~al.}(2019{\natexlab{b}})\citenamefont {Hendrickx}, \citenamefont
  {Tagliaferri}, \citenamefont {Kouwenhoven}, \citenamefont {Li}, \citenamefont
  {Franke}, \citenamefont {Sammak}, \citenamefont {Brinkman}, \citenamefont
  {Scappucci},\ and\ \citenamefont {Veldhorst}}]{delft:ge}%
  \BibitemOpen
  \bibfield  {author} {\bibinfo {author} {\bibfnamefont {N.~W.}\ \bibnamefont
  {Hendrickx}}, \bibinfo {author} {\bibfnamefont {M.~L.~V.}\ \bibnamefont
  {Tagliaferri}}, \bibinfo {author} {\bibfnamefont {M.}~\bibnamefont
  {Kouwenhoven}}, \bibinfo {author} {\bibfnamefont {R.}~\bibnamefont {Li}},
  \bibinfo {author} {\bibfnamefont {D.~P.}\ \bibnamefont {Franke}}, \bibinfo
  {author} {\bibfnamefont {A.}~\bibnamefont {Sammak}}, \bibinfo {author}
  {\bibfnamefont {A.}~\bibnamefont {Brinkman}}, \bibinfo {author}
  {\bibfnamefont {G.}~\bibnamefont {Scappucci}}, \ and\ \bibinfo {author}
  {\bibfnamefont {M.}~\bibnamefont {Veldhorst}},\ }\href {\doibase
  10.1103/PhysRevB.99.075435} {\bibfield  {journal} {\bibinfo  {journal} {Phys.
  Rev. B}\ }\textbf {\bibinfo {volume} {99}},\ \bibinfo {pages} {075435}
  (\bibinfo {year} {2019}{\natexlab{b}})}\BibitemShut {NoStop}%
\bibitem [{\citenamefont {Blamire}\ \emph {et~al.}(1991)\citenamefont
  {Blamire}, \citenamefont {Kirk}, \citenamefont {Evetts},\ and\ \citenamefont
  {Klapwijk}}]{klapwijk:al-nb}%
  \BibitemOpen
  \bibfield  {author} {\bibinfo {author} {\bibfnamefont {M.~G.}\ \bibnamefont
  {Blamire}}, \bibinfo {author} {\bibfnamefont {E.~C.~G.}\ \bibnamefont
  {Kirk}}, \bibinfo {author} {\bibfnamefont {J.~E.}\ \bibnamefont {Evetts}}, \
  and\ \bibinfo {author} {\bibfnamefont {T.~M.}\ \bibnamefont {Klapwijk}},\
  }\href {\doibase 10.1103/PhysRevLett.66.220} {\bibfield  {journal} {\bibinfo
  {journal} {Phys. Rev. Lett.}\ }\textbf {\bibinfo {volume} {66}},\ \bibinfo
  {pages} {220} (\bibinfo {year} {1991})}\BibitemShut {NoStop}%
\bibitem [{\citenamefont {Beenakker}(1991)}]{Beenakker1991a}%
  \BibitemOpen
  \bibfield  {author} {\bibinfo {author} {\bibfnamefont {C.~W.~J.}\
  \bibnamefont {Beenakker}},\ }\href {\doibase 10.1103/PhysRevLett.67.3836}
  {\bibfield  {journal} {\bibinfo  {journal} {Phys. Rev. Lett.}\ }\textbf
  {\bibinfo {volume} {67}},\ \bibinfo {pages} {3836} (\bibinfo {year}
  {1991})}\BibitemShut {NoStop}%
\bibitem [{\citenamefont {Klapwijk}\ \emph {et~al.}(1982)\citenamefont
  {Klapwijk}, \citenamefont {Blonder},\ and\ \citenamefont
  {Tinkham}}]{Klapwijk1982}%
  \BibitemOpen
  \bibfield  {author} {\bibinfo {author} {\bibfnamefont {T.}~\bibnamefont
  {Klapwijk}}, \bibinfo {author} {\bibfnamefont {G.}~\bibnamefont {Blonder}}, \
  and\ \bibinfo {author} {\bibfnamefont {M.}~\bibnamefont {Tinkham}},\ }\href
  {\doibase https://doi.org/10.1016/0378-4363(82)90189-9} {\bibfield  {journal}
  {\bibinfo  {journal} {Physica B+C}\ }\textbf {\bibinfo {volume} {109-110}},\
  \bibinfo {pages} {1657} (\bibinfo {year} {1982})}\BibitemShut {NoStop}%
\bibitem [{\citenamefont {Matthias}\ \emph {et~al.}(1963)\citenamefont
  {Matthias}, \citenamefont {Geballe},\ and\ \citenamefont
  {Compton}}]{Matthias1963}%
  \BibitemOpen
  \bibfield  {author} {\bibinfo {author} {\bibfnamefont {B.~T.}\ \bibnamefont
  {Matthias}}, \bibinfo {author} {\bibfnamefont {T.~H.}\ \bibnamefont
  {Geballe}}, \ and\ \bibinfo {author} {\bibfnamefont {V.~B.}\ \bibnamefont
  {Compton}},\ }\href {\doibase 10.1103/RevModPhys.35.1} {\bibfield  {journal}
  {\bibinfo  {journal} {Rev. Mod. Phys.}\ }\textbf {\bibinfo {volume} {35}},\
  \bibinfo {pages} {1} (\bibinfo {year} {1963})}\BibitemShut {NoStop}%
\bibitem [{\citenamefont {Drachmann}\ \emph {et~al.}(2017)\citenamefont
  {Drachmann}, \citenamefont {Suominen}, \citenamefont {Kjaergaard},
  \citenamefont {Shojaei}, \citenamefont {Palmstrøm}, \citenamefont {Marcus},\
  and\ \citenamefont {Nichele}}]{Drachmann2017}%
  \BibitemOpen
  \bibfield  {author} {\bibinfo {author} {\bibfnamefont {A.~C.~C.}\
  \bibnamefont {Drachmann}}, \bibinfo {author} {\bibfnamefont {H.~J.}\
  \bibnamefont {Suominen}}, \bibinfo {author} {\bibfnamefont {M.}~\bibnamefont
  {Kjaergaard}}, \bibinfo {author} {\bibfnamefont {B.}~\bibnamefont {Shojaei}},
  \bibinfo {author} {\bibfnamefont {C.~J.}\ \bibnamefont {Palmstrøm}},
  \bibinfo {author} {\bibfnamefont {C.~M.}\ \bibnamefont {Marcus}}, \ and\
  \bibinfo {author} {\bibfnamefont {F.}~\bibnamefont {Nichele}},\ }\href
  {\doibase 10.1021/acs.nanolett.6b04964} {\bibfield  {journal} {\bibinfo
  {journal} {Nano Letters}\ }\textbf {\bibinfo {volume} {17}},\ \bibinfo
  {pages} {1200} (\bibinfo {year} {2017})}\BibitemShut {NoStop}%
\bibitem [{\citenamefont {Kjaergaard}\ \emph {et~al.}(2017)\citenamefont
  {Kjaergaard}, \citenamefont {Suominen}, \citenamefont {Nowak}, \citenamefont
  {Akhmerov}, \citenamefont {Shabani}, \citenamefont {Palmstr\o{}m},
  \citenamefont {Nichele},\ and\ \citenamefont {Marcus}}]{Kjaergaard2017}%
  \BibitemOpen
  \bibfield  {author} {\bibinfo {author} {\bibfnamefont {M.}~\bibnamefont
  {Kjaergaard}}, \bibinfo {author} {\bibfnamefont {H.~J.}\ \bibnamefont
  {Suominen}}, \bibinfo {author} {\bibfnamefont {M.~P.}\ \bibnamefont {Nowak}},
  \bibinfo {author} {\bibfnamefont {A.~R.}\ \bibnamefont {Akhmerov}}, \bibinfo
  {author} {\bibfnamefont {J.}~\bibnamefont {Shabani}}, \bibinfo {author}
  {\bibfnamefont {C.~J.}\ \bibnamefont {Palmstr\o{}m}}, \bibinfo {author}
  {\bibfnamefont {F.}~\bibnamefont {Nichele}}, \ and\ \bibinfo {author}
  {\bibfnamefont {C.~M.}\ \bibnamefont {Marcus}},\ }\href {\doibase
  10.1103/PhysRevApplied.7.034029} {\bibfield  {journal} {\bibinfo  {journal}
  {Phys. Rev. Applied}\ }\textbf {\bibinfo {volume} {7}},\ \bibinfo {pages}
  {034029} (\bibinfo {year} {2017})}\BibitemShut {NoStop}%
\bibitem [{\citenamefont {Averin}\ and\ \citenamefont
  {Bardas}(1995)}]{Averin1995}%
  \BibitemOpen
  \bibfield  {author} {\bibinfo {author} {\bibfnamefont {D.}~\bibnamefont
  {Averin}}\ and\ \bibinfo {author} {\bibfnamefont {A.}~\bibnamefont
  {Bardas}},\ }\href {\doibase 10.1103/PhysRevLett.75.1831} {\bibfield
  {journal} {\bibinfo  {journal} {Phys. Rev. Lett.}\ }\textbf {\bibinfo
  {volume} {75}},\ \bibinfo {pages} {1831} (\bibinfo {year}
  {1995})}\BibitemShut {NoStop}%
\bibitem [{\citenamefont {Flensberg}\ \emph {et~al.}(1988)\citenamefont
  {Flensberg}, \citenamefont {Hansen},\ and\ \citenamefont
  {Octavio}}]{flensberg1988}%
  \BibitemOpen
  \bibfield  {author} {\bibinfo {author} {\bibfnamefont {K.}~\bibnamefont
  {Flensberg}}, \bibinfo {author} {\bibfnamefont {J.~B.}\ \bibnamefont
  {Hansen}}, \ and\ \bibinfo {author} {\bibfnamefont {M.}~\bibnamefont
  {Octavio}},\ }\href {\doibase 10.1103/PhysRevB.38.8707} {\bibfield  {journal}
  {\bibinfo  {journal} {Phys. Rev. B}\ }\textbf {\bibinfo {volume} {38}},\
  \bibinfo {pages} {8707} (\bibinfo {year} {1988})}\BibitemShut {NoStop}%
\bibitem [{\citenamefont {Octavio}\ \emph {et~al.}(1983)\citenamefont
  {Octavio}, \citenamefont {Tinkham}, \citenamefont {Blonder},\ and\
  \citenamefont {Klapwijk}}]{octavio1983}%
  \BibitemOpen
  \bibfield  {author} {\bibinfo {author} {\bibfnamefont {M.}~\bibnamefont
  {Octavio}}, \bibinfo {author} {\bibfnamefont {M.}~\bibnamefont {Tinkham}},
  \bibinfo {author} {\bibfnamefont {G.~E.}\ \bibnamefont {Blonder}}, \ and\
  \bibinfo {author} {\bibfnamefont {T.~M.}\ \bibnamefont {Klapwijk}},\ }\href
  {\doibase 10.1103/PhysRevB.27.6739} {\bibfield  {journal} {\bibinfo
  {journal} {Phys. Rev. B}\ }\textbf {\bibinfo {volume} {27}},\ \bibinfo
  {pages} {6739} (\bibinfo {year} {1983})}\BibitemShut {NoStop}%
\bibitem [{\citenamefont {Kresin}(1986)}]{kresin:thouless}%
  \BibitemOpen
  \bibfield  {author} {\bibinfo {author} {\bibfnamefont {V.~Z.}\ \bibnamefont
  {Kresin}},\ }\href {\doibase 10.1103/PhysRevB.34.7587} {\bibfield  {journal}
  {\bibinfo  {journal} {Phys. Rev. B}\ }\textbf {\bibinfo {volume} {34}},\
  \bibinfo {pages} {7587} (\bibinfo {year} {1986})}\BibitemShut {NoStop}%
\bibitem [{Not()}]{Note1}%
  \BibitemOpen
  \href@noop {} {}\bibinfo {note} {Being not strictly in the short-junction
  limit could also provide an explanation for the rather low value we found for
  $I_{\rm C}R_{\rm N}$~\cite{schon:sns}.}\BibitemShut {Stop}%
\bibitem [{\citenamefont {Beenakker}\ and\ \citenamefont {van
  Houten}(1991)}]{Beenakker1991}%
  \BibitemOpen
  \bibfield  {author} {\bibinfo {author} {\bibfnamefont {C.~W.~J.}\
  \bibnamefont {Beenakker}}\ and\ \bibinfo {author} {\bibfnamefont
  {H.}~\bibnamefont {van Houten}},\ }\href {\doibase
  10.1103/PhysRevLett.66.3056} {\bibfield  {journal} {\bibinfo  {journal}
  {Phys. Rev. Lett.}\ }\textbf {\bibinfo {volume} {66}},\ \bibinfo {pages}
  {3056} (\bibinfo {year} {1991})}\BibitemShut {NoStop}%
\bibitem [{\citenamefont {Dorokhov}(1984)}]{Dorokhov1984}%
  \BibitemOpen
  \bibfield  {author} {\bibinfo {author} {\bibfnamefont {O.}~\bibnamefont
  {Dorokhov}},\ }\href {\doibase https://doi.org/10.1016/0038-1098(84)90117-0}
  {\bibfield  {journal} {\bibinfo  {journal} {Solid State Communications}\
  }\textbf {\bibinfo {volume} {51}},\ \bibinfo {pages} {381} (\bibinfo {year}
  {1984})}\BibitemShut {NoStop}%
\bibitem [{\citenamefont {Nazarov}(1994)}]{Nazarov1994}%
  \BibitemOpen
  \bibfield  {author} {\bibinfo {author} {\bibfnamefont {Y.~V.}\ \bibnamefont
  {Nazarov}},\ }\href {\doibase 10.1103/PhysRevLett.73.134} {\bibfield
  {journal} {\bibinfo  {journal} {Phys. Rev. Lett.}\ }\textbf {\bibinfo
  {volume} {73}},\ \bibinfo {pages} {134} (\bibinfo {year} {1994})}\BibitemShut
  {NoStop}%
\bibitem [{\citenamefont {Golubov}\ \emph {et~al.}(2004)\citenamefont
  {Golubov}, \citenamefont {Kupriyanov},\ and\ \citenamefont
  {Il'ichev}}]{Golubov2004}%
  \BibitemOpen
  \bibfield  {author} {\bibinfo {author} {\bibfnamefont {A.~A.}\ \bibnamefont
  {Golubov}}, \bibinfo {author} {\bibfnamefont {M.~Y.}\ \bibnamefont
  {Kupriyanov}}, \ and\ \bibinfo {author} {\bibfnamefont {E.}~\bibnamefont
  {Il'ichev}},\ }\href {\doibase 10.1103/RevModPhys.76.411} {\bibfield
  {journal} {\bibinfo  {journal} {Rev. Mod. Phys.}\ }\textbf {\bibinfo {volume}
  {76}},\ \bibinfo {pages} {411} (\bibinfo {year} {2004})}\BibitemShut
  {NoStop}%
\bibitem [{\citenamefont {{Parks}}(1968)}]{parks1968}%
  \BibitemOpen
  \bibfield  {author} {\bibinfo {author} {\bibfnamefont {R.~D.}\ \bibnamefont
  {{Parks}}},\ }\href {\doibase 10.1063/1.1656600} {\bibfield  {journal}
  {\bibinfo  {journal} {Journal of Applied Physics}\ }\textbf {\bibinfo
  {volume} {39}},\ \bibinfo {pages} {2515} (\bibinfo {year}
  {1968})}\BibitemShut {NoStop}%
\bibitem [{\citenamefont {Rasmussen}\ \emph {et~al.}(2016)\citenamefont
  {Rasmussen}, \citenamefont {Danon}, \citenamefont {Suominen}, \citenamefont
  {Nichele}, \citenamefont {Kjaergaard},\ and\ \citenamefont
  {Flensberg}}]{rasmussen2016}%
  \BibitemOpen
  \bibfield  {author} {\bibinfo {author} {\bibfnamefont {A.}~\bibnamefont
  {Rasmussen}}, \bibinfo {author} {\bibfnamefont {J.}~\bibnamefont {Danon}},
  \bibinfo {author} {\bibfnamefont {H.}~\bibnamefont {Suominen}}, \bibinfo
  {author} {\bibfnamefont {F.}~\bibnamefont {Nichele}}, \bibinfo {author}
  {\bibfnamefont {M.}~\bibnamefont {Kjaergaard}}, \ and\ \bibinfo {author}
  {\bibfnamefont {K.}~\bibnamefont {Flensberg}},\ }\href {\doibase
  10.1103/PhysRevB.93.155406} {\bibfield  {journal} {\bibinfo  {journal} {Phys.
  Rev. B}\ }\textbf {\bibinfo {volume} {93}},\ \bibinfo {pages} {155406}
  (\bibinfo {year} {2016})}\BibitemShut {NoStop}%
\bibitem [{\citenamefont {{Khukhareva}}(1963)}]{Khukhareva1963}%
  \BibitemOpen
  \bibfield  {author} {\bibinfo {author} {\bibfnamefont {I.~S.}\ \bibnamefont
  {{Khukhareva}}},\ }\href
  {https://ui.adsabs.harvard.edu/abs/1963JETP...16..828K} {\bibfield  {journal}
  {\bibinfo  {journal} {Soviet Journal of Experimental and Theoretical
  Physics}\ }\textbf {\bibinfo {volume} {16}},\ \bibinfo {pages} {828}
  (\bibinfo {year} {1963})}\BibitemShut {NoStop}%
\bibitem [{\citenamefont {Nichele}\ \emph {et~al.}(2020)\citenamefont
  {Nichele}, \citenamefont {Portol\'es}, \citenamefont {Fornieri},
  \citenamefont {Whiticar}, \citenamefont {Drachmann}, \citenamefont {Gronin},
  \citenamefont {Wang}, \citenamefont {Gardner}, \citenamefont {Thomas},
  \citenamefont {Hatke}, \citenamefont {Manfra},\ and\ \citenamefont
  {Marcus}}]{Nichele2020}%
  \BibitemOpen
  \bibfield  {author} {\bibinfo {author} {\bibfnamefont {F.}~\bibnamefont
  {Nichele}}, \bibinfo {author} {\bibfnamefont {E.}~\bibnamefont {Portol\'es}},
  \bibinfo {author} {\bibfnamefont {A.}~\bibnamefont {Fornieri}}, \bibinfo
  {author} {\bibfnamefont {A.~M.}\ \bibnamefont {Whiticar}}, \bibinfo {author}
  {\bibfnamefont {A.~C.~C.}\ \bibnamefont {Drachmann}}, \bibinfo {author}
  {\bibfnamefont {S.}~\bibnamefont {Gronin}}, \bibinfo {author} {\bibfnamefont
  {T.}~\bibnamefont {Wang}}, \bibinfo {author} {\bibfnamefont {G.~C.}\
  \bibnamefont {Gardner}}, \bibinfo {author} {\bibfnamefont {C.}~\bibnamefont
  {Thomas}}, \bibinfo {author} {\bibfnamefont {A.~T.}\ \bibnamefont {Hatke}},
  \bibinfo {author} {\bibfnamefont {M.~J.}\ \bibnamefont {Manfra}}, \ and\
  \bibinfo {author} {\bibfnamefont {C.~M.}\ \bibnamefont {Marcus}},\ }\href
  {\doibase 10.1103/PhysRevLett.124.226801} {\bibfield  {journal} {\bibinfo
  {journal} {Phys. Rev. Lett.}\ }\textbf {\bibinfo {volume} {124}},\ \bibinfo
  {pages} {226801} (\bibinfo {year} {2020})}\BibitemShut {NoStop}%
\bibitem [{\citenamefont {Szombati}\ \emph {et~al.}(2016)\citenamefont
  {Szombati}, \citenamefont {Nadj-Perge}, \citenamefont {Car}, \citenamefont
  {Plissard}, \citenamefont {Bakkers},\ and\ \citenamefont
  {Kouwenhoven}}]{Szombati2016}%
  \BibitemOpen
  \bibfield  {author} {\bibinfo {author} {\bibfnamefont {D.~B.}\ \bibnamefont
  {Szombati}}, \bibinfo {author} {\bibfnamefont {S.}~\bibnamefont
  {Nadj-Perge}}, \bibinfo {author} {\bibfnamefont {D.}~\bibnamefont {Car}},
  \bibinfo {author} {\bibfnamefont {S.~R.}\ \bibnamefont {Plissard}}, \bibinfo
  {author} {\bibfnamefont {E.~P. A.~M.}\ \bibnamefont {Bakkers}}, \ and\
  \bibinfo {author} {\bibfnamefont {L.~P.}\ \bibnamefont {Kouwenhoven}},\
  }\href {\doibase 10.1038/nphys3742} {\bibfield  {journal} {\bibinfo
  {journal} {Nature Physics}\ }\textbf {\bibinfo {volume} {12}},\ \bibinfo
  {pages} {568} (\bibinfo {year} {2016})}\BibitemShut {NoStop}%
\bibitem [{\citenamefont {Mayer}\ \emph {et~al.}(2020)\citenamefont {Mayer},
  \citenamefont {Dartiailh}, \citenamefont {Yuan}, \citenamefont
  {Wickramasinghe}, \citenamefont {Rossi},\ and\ \citenamefont
  {Shabani}}]{Mayer2020}%
  \BibitemOpen
  \bibfield  {author} {\bibinfo {author} {\bibfnamefont {W.}~\bibnamefont
  {Mayer}}, \bibinfo {author} {\bibfnamefont {M.~C.}\ \bibnamefont
  {Dartiailh}}, \bibinfo {author} {\bibfnamefont {J.}~\bibnamefont {Yuan}},
  \bibinfo {author} {\bibfnamefont {K.~S.}\ \bibnamefont {Wickramasinghe}},
  \bibinfo {author} {\bibfnamefont {E.}~\bibnamefont {Rossi}}, \ and\ \bibinfo
  {author} {\bibfnamefont {J.}~\bibnamefont {Shabani}},\ }\href
  {https://doi.org/10.1038/s41467-019-14094-1} {\bibfield  {journal} {\bibinfo
  {journal} {Nature Communications}\ }\textbf {\bibinfo {volume} {11}},\
  \bibinfo {pages} {212} (\bibinfo {year} {2020})}\BibitemShut {NoStop}%
\bibitem [{\citenamefont {Assouline}\ \emph {et~al.}(2019)\citenamefont
  {Assouline}, \citenamefont {Feuillet-Palma}, \citenamefont {Bergeal},
  \citenamefont {Zhang}, \citenamefont {Mottaghizadeh}, \citenamefont
  {Zimmers}, \citenamefont {Lhuillier}, \citenamefont {Eddrie}, \citenamefont
  {Atkinson}, \citenamefont {Aprili},\ and\ \citenamefont
  {Aubin}}]{Assouline2019}%
  \BibitemOpen
  \bibfield  {author} {\bibinfo {author} {\bibfnamefont {A.}~\bibnamefont
  {Assouline}}, \bibinfo {author} {\bibfnamefont {C.}~\bibnamefont
  {Feuillet-Palma}}, \bibinfo {author} {\bibfnamefont {N.}~\bibnamefont
  {Bergeal}}, \bibinfo {author} {\bibfnamefont {T.}~\bibnamefont {Zhang}},
  \bibinfo {author} {\bibfnamefont {A.}~\bibnamefont {Mottaghizadeh}}, \bibinfo
  {author} {\bibfnamefont {A.}~\bibnamefont {Zimmers}}, \bibinfo {author}
  {\bibfnamefont {E.}~\bibnamefont {Lhuillier}}, \bibinfo {author}
  {\bibfnamefont {M.}~\bibnamefont {Eddrie}}, \bibinfo {author} {\bibfnamefont
  {P.}~\bibnamefont {Atkinson}}, \bibinfo {author} {\bibfnamefont
  {M.}~\bibnamefont {Aprili}}, \ and\ \bibinfo {author} {\bibfnamefont
  {H.}~\bibnamefont {Aubin}},\ }\href
  {https://doi.org/10.1038/s41467-018-08022-y} {\bibfield  {journal} {\bibinfo
  {journal} {Nature Communications}\ }\textbf {\bibinfo {volume} {10}},\
  \bibinfo {pages} {126} (\bibinfo {year} {2019})}\BibitemShut {NoStop}%
\bibitem [{\citenamefont {{Kim}}\ \emph {et~al.}(2015)\citenamefont {{Kim}},
  \citenamefont {{Kim}}, \citenamefont {{Kim}}, \citenamefont {{Shin}},
  \citenamefont {{Park}}, \citenamefont {{Saraswat}}, \citenamefont {{Cho}},\
  and\ \citenamefont {{Yu}}}]{7116476}%
  \BibitemOpen
  \bibfield  {author} {\bibinfo {author} {\bibfnamefont {G.}~\bibnamefont
  {{Kim}}}, \bibinfo {author} {\bibfnamefont {S.}~\bibnamefont {{Kim}}},
  \bibinfo {author} {\bibfnamefont {J.}~\bibnamefont {{Kim}}}, \bibinfo
  {author} {\bibfnamefont {C.}~\bibnamefont {{Shin}}}, \bibinfo {author}
  {\bibfnamefont {J.}~\bibnamefont {{Park}}}, \bibinfo {author} {\bibfnamefont
  {K.~C.}\ \bibnamefont {{Saraswat}}}, \bibinfo {author} {\bibfnamefont
  {B.~J.}\ \bibnamefont {{Cho}}}, \ and\ \bibinfo {author} {\bibfnamefont
  {H.}~\bibnamefont {{Yu}}},\ }\href {\doibase 10.1109/LED.2015.2440434}
  {\bibfield  {journal} {\bibinfo  {journal} {IEEE Electron Device Letters}\
  }\textbf {\bibinfo {volume} {36}},\ \bibinfo {pages} {745} (\bibinfo {year}
  {2015})}\BibitemShut {NoStop}%
\bibitem [{\citenamefont {Dubos}\ \emph {et~al.}(2001)\citenamefont {Dubos},
  \citenamefont {Courtois}, \citenamefont {Pannetier}, \citenamefont {Wilhelm},
  \citenamefont {Zaikin},\ and\ \citenamefont {Sch\"on}}]{schon:sns}%
  \BibitemOpen
  \bibfield  {author} {\bibinfo {author} {\bibfnamefont {P.}~\bibnamefont
  {Dubos}}, \bibinfo {author} {\bibfnamefont {H.}~\bibnamefont {Courtois}},
  \bibinfo {author} {\bibfnamefont {B.}~\bibnamefont {Pannetier}}, \bibinfo
  {author} {\bibfnamefont {F.~K.}\ \bibnamefont {Wilhelm}}, \bibinfo {author}
  {\bibfnamefont {A.~D.}\ \bibnamefont {Zaikin}}, \ and\ \bibinfo {author}
  {\bibfnamefont {G.}~\bibnamefont {Sch\"on}},\ }\href {\doibase
  10.1103/PhysRevB.63.064502} {\bibfield  {journal} {\bibinfo  {journal} {Phys.
  Rev. B}\ }\textbf {\bibinfo {volume} {63}},\ \bibinfo {pages} {064502}
  (\bibinfo {year} {2001})}\BibitemShut {NoStop}%
\end{thebibliography}%

\end{document}